# Exoplanet secondary atmosphere loss and revival

Edwin S. Kite[1,*] and Megan Barnett[1].
1. University of Chicago, Chicago, IL. *Corresponding author. kite@uchicago.edu

**Abstract.**

The next step on the path toward another Earth is to find atmospheres similar to those of Earth and Venus – high-molecular-weight (secondary) atmospheres – on rocky exoplanets. Many rocky exoplanets are born with thick (> 10 kbar) $H_2$-dominated atmospheres but subsequently lose their $H_2$; this process has no known Solar System analog. We study the consequences of early loss of a thick $H_2$ atmosphere for subsequent occurrence of a high-molecular-weight atmosphere using a simple model of atmosphere evolution (including atmosphere loss to space, magma ocean crystallization, and volcanic outgassing). We also calculate atmosphere survival for rocky worlds that start with no $H_2$. Our results imply that most rocky exoplanets orbiting closer to their star than the Habitable Zone that were formed with thick $H_2$-dominated atmospheres lack high-molecular-weight atmospheres today. During early magma ocean crystallization, high-molecular-weight species usually do not form long-lived high-molecular-weight atmospheres; instead they are lost to space alongside $H_2$. This early volatile depletion also makes it more difficult for later volcanic outgassing to revive the atmosphere. However, atmospheres should persist on worlds that start with abundant volatiles (for example, waterworlds). Our results imply that in order to find high–molecular-weight atmospheres on warm exoplanets orbiting M-stars, we should target worlds that formed $H_2$-poor, that have anomalously large radii, or that orbit less active stars.

**Significance statement.**

Earth and Venus have significant atmospheres, but Mercury does not. Thousands of exoplanets are known, but we know almost nothing about the rocky exoplanet atmospheres. Many rocky exoplanets were formed by a sub-Neptune to super-Earth conversion process during which planets lose most of their $H_2$-rich (primary) atmospheres and are reduced in volume by a factor of >2. Does such a gas-rich adolescence increase or decrease the likelihood that super-Earths will subsequently exhibit a $H_2$-poor (secondary) atmosphere? We show that secondary atmospheres exsolved from the magma ocean are unlikely to be retained by super-Earths, but it is possible for volcanic outgassing to revive super-Earth atmospheres. For M-dwarf planetary systems, super-Earths that have atmospheres close to the star likely were formed with abundant volatiles.

**Introduction.**

The Solar System has three planets – Earth, Venus, and Mars – that have atmospheres derived from $H_2$-free solids (secondary atmospheres), and four giant planets whose atmospheres are derived from protoplanetary nebula gas ($H_2$-dominated primary atmospheres) (Atreya et al. 1989, Marty et al. 2016). However, this clean separation in process and outcome is apparently unrepresentative of the known exoplanets.

The two most common types of known exoplanet, the rocky super-Earth-sized planets (planet radius $R_{pl}$ < 1.6 $R_⊕$, where "⊕" is the Earth symbol, planet density $\rho_{pl}$ > 4 g/cc; "super-Earths") and the gas-shrouded sub-Neptunes ($R_{pl}$ = 2 - 3 $R_⊕$), are divided by a valley in (planet radius)-(orbital period) space in which planets are less common (Fig. 1) (Fulton et al. 2017). The radius valley can be understood if, and only if, a substantial fraction of planets that are born with thick (>10 kbar) $H_2$-dominated primary atmospheres lose those atmospheres and shrink in radius to become rocky super-Earths (Van Eylen et al. 2018). This radius-shrinking process, which carves out the radius valley, may be the way that most super-Earths form. There is no evidence that Earth and Venus underwent this process, and primary atmospheres are not thought to have contributed to the origin of major volatile elements in Earth (Dauphas & Morbidelli 2014). Both Venus and Earth have secondary volatile envelopes (composed of solid-derived volatiles, including $H_2O$ and $CO_2$) that are much less massive than the atmospheres of sub-Neptunes. Large surface reservoirs of $H_2O$, C species, and N species are essential to life on Earth and to Earth's habitable climate (Bergin et al. 2015, Catling & Kasting 2017).

Does forming with a thick primary atmosphere (sub-Neptune) but ending up as a rocky super-Earth favor the rocky planet ending up with a secondary atmosphere? Do primary atmospheres, in dying, shield high-molecular-weight species from loss during the early era of atmosphere-stripping impacts and intense stellar activity, allowing those constituents to later form a secondary atmosphere? Or does the light ($H_2$-dominated) and transient primary atmosphere drag away the higher-molecular weight species? Getting physical insight into the transition from primary to secondary atmospheres is particularly important for rocky exoplanets that are too hot for life. Hot rocky exoplanets are the highest signal/noise rocky targets for upcoming missions such as James Webb Space Telescope (JWST) (Kempton et al. 2018) and so will be the most useful for checking our understanding of this atmospheric transition process.

Secondary atmospheres are central to the exoplanet exploration strategy (National Academy of Sciences 2018, Zahnle & Catling 2017). Previous work on the hypothesis that primary atmospheres played a role in forming secondary atmospheres includes that of Eucken in the 1940s (Turekian & Clark 1969), Urey (1951), Cameron and coworkers (e.g. Slattery et al. 1980), Sasaki (1990), and Ozima & Zahnle (1993). Secondary atmosphere formation on exoplanets in the absence of a primary atmosphere has been investigated by (e.g.) Elkins-Tanton & Seager (2008), Dorn et al. (2018), and references therein.

**Gas survival during planetary volume reduction.**

During sub-Neptune to super-Earth conversion, we suppose the planet contains both nebular-derived $H_2$ and also high-$\mu$ species (derived from the planet-forming solid materials) that could form a secondary atmosphere – if retained. Retention of volatiles is controlled by atmospheric loss and atmosphere-interior exchange (Fig. 2). For both sub-Neptunes and super-Earths, silicates (both magma and solid rock) apparently make up most of the planet's mass (e.g. Dai et al. 2019, Owen and Wu 2017).

Atmosphere loss is more difficult for high-molecular-weight volatiles than for $H_2$ because higher-molecular-weight volatiles such as $H_2O$ are more easily shielded within the silicate interior and are also more resistant to escape.

This can be understood as follows. First, a basic control on atmosphere loss is the ratio, $\lambda$, of gravitational binding energy to thermal energy:

$$\lambda = (\mu / k\, T_{ua})\, (G\, M_{pl} / (R + z))$$
$$\approx 2.3\, \mu\, (10^4\,\text{K} / T_{ua})(M_{pl} / 6\, M_\oplus)/((R + z) / 2\, R_\oplus) \qquad (1)$$

where $\mu$ is molecular mass, $k$ is Boltzmann's constant, $T_{ua}$ is upper-atmosphere temperature, $G$ is the gravitational constant, $M_{pl}$ is planet mass, $R$ is the radius of the silicate planet, and $z$ is atmosphere thickness. When $\lambda \lesssim 2$, the upper atmosphere flows out to space at a rate potentially limited only by the energy available from upper-atmosphere absorption of light from the star (Watson et al. 1981). If the upper atmosphere absorbs 100 W/m² of light from the star, the upper limit on loss is ~1000 bars/Myr. This hydrodynamic outflow ejects the tenuous upper atmosphere at high speed, but the dense lower atmosphere remains close to hydrostatic equilibrium. When $\lambda > 10$, the hydrodynamic outflow shuts down. For primary atmospheres, the mean molecular weight $\mu_{avg} \approx 2$ Da ($H_2$) while for secondary atmospheres, $\mu_{avg}$ is $\gtrsim 10\times$ higher favoring secondary atmosphere retention. Moreover, many secondary-atmosphere constituents (e.g $CO_2$) are much more effective coolants than $H_2$, so for secondary atmospheres $T_{ua}$ is lower (e.g. Kulikov et al. 2007, Johnstone et al. 2018). Atmospheric thickness $z$ is also smaller for high-$\mu_{avg}$ atmospheres due to their smaller scale height, raising $\lambda$. For all atmospheres, proximity to the star increases $T_{ua}$ (lowering $\lambda$) and also increases the upper atmosphere outflow rate when the $\lambda \lesssim 2$ condition is satisfied. Moreover, closer to the star impacts occur at higher velocities that are more erosive (Kegerreis et al. 2020). Thus we expect massive worlds far from the star to retain atmospheres and low-mass worlds closer to the star to lose them (Zahnle & Catling 2017). This expectation, while physically valid, offers little guidance as to whether or not super-Earths will have atmospheres. To go further, we need to consider the effect of evolving atmospheric composition on loss rate (Eqn. 1) and also track the shielding of volatiles within silicates (magma oceans and solid rock) (Fig. 2).

The atmospheres of young sub-Neptunes are underlain by magma oceans (e.g. Vazan et al. 2018; Fig. S1). Equilibrium partitioning of a volatile $i$ between the atmosphere (where it is vulnerable to escape) and the magma ocean (where it is shielded from escape) is given by the following:

$$c_i = e_i + \quad p_i\, s_i\, m_{magma} + \text{other reservoirs}$$
$$= e_i + [\, (e_i / A_{mai})\, g_{mai}\, (\mu_{avg} / \mu_i)\,]\, s_i\, m_{magma} + \text{other reservoirs} \qquad (2)$$

where $c_i$ is the total inventory of the volatile, $e_i$ is the mass in the atmosphere, $p_i$ is the partial pressure of $i$ at the magma-atmosphere interface, $A_{mai}$ (and $g_{mai}$) are the area of (and gravitational acceleration at) the magma-atmosphere interface, $\mu_{avg}$ is the mean $\mu$ of the atmosphere, $s_i$ is the solubility coefficient (mass fraction pascal$^{-1}$) of the volatile in the magma, and $m_{magma}$ is the mass of liquid in the magma ocean. Solubility in magma is higher for many secondary-atmosphere constituents (e.g. $CH_4$, $H_2S$, $H_2O$) than for $H_2$. When the atmosphere contains both $H_2$ and a more-soluble, high-$\mu$ constituent, the $H_2$ takes the brunt of atmosphere loss processes that might otherwise remove the high-$\mu$ species. Differential solubility also protects the more-soluble species from impact shock (Genda & Abe 2005). Other reservoirs include volatiles stored in the crystal phase. This is an unimportant reservoir early on when the planet has a massive magma ocean, but becomes important later, because these volatiles can be released by volcanic degassing. We neglect volatiles that are stored in the iron core because they do not contribute to the observable atmosphere.

Atmospheric loss typically drives magma ocean crystallization (Fig. S1) (except for planets that either orbit uncommonly close to their star, or undergo strong tidal heating, bare-rock planets do not have a global molten-rock layer). The greenhouse effect from the atmosphere keeps the magma ocean liquid for longer, which delays partitioning of the volatile into crystals. This partitioning is described by the following:

$$X_{xtl} = D_i X_i \qquad (3)$$

where $X_{xtl}$ is the concentration of $i$ in the crystal phase, $D_i$ is a solid-melt distribution coefficient / partition coefficient, and $X_i$ is the concentration of $i$ in the magma. Crystallization enriches the residual melt in volatiles (i.e. $D_i << 1$; e.g. Elkins-Tanton 2008).

One last effect can aid retention of high-molecular-weight volatiles. If outflow to space is slow, then the high-$\mu$ species can sink back toward the planet (diffusive separation driven by buoyancy) (e.g. Hunten et al. 1987, Hu et al. 2015; Supplementary Information 1g).

Eqns. (1)-(3) open two paths from a primary to a secondary atmosphere. A secondary atmosphere can be exsolved as the magma ocean crystallizes. Alternatively, the atmosphere can be revived by volcanic degassing long after the mantle has almost entirely solidified (Fig. 2). We explore these paths below.

**Model of the transition from primary to secondary atmospheres.**

We model the effect of the loss of a thick $H_2$-dominated atmosphere on the retention (or loss) of a hypothetical volatile, $s$. Species $s$ has molecular mass 30 Da, which is within a factor of <1.5 of the molecular mass of all known major secondary-atmosphere constituents. In the uppermost atmosphere, species $s$ is modeled to split into fragments of mass 15 Da for the purpose of calculating whether or not $s$ is effectively entrained by escaping H. We assume that reactions between $s$ and $H_2$ do not affect the mass inventory of either species; in effect, $s$ is chemically inert. We assume that deviations from thermochemical equilibrium driven by photochemistry in the uppermost atmosphere are reset to thermochemical equilibrium at the high temperature and pressure of the magma-atmosphere interface. While idealized, the model includes many effects that have not previously been incorporated into a model of atmosphere loss, such as a realistic rock melting curve, differential solubility effects, etc. (Supplementary Information).

Planet equilibrium temperature $T_{eq}$ (in K), for zero planet albedo, is given by the following:

$$T_{eq} = 278 \, (F/F_\oplus)^{1/4} \qquad (4)$$

where $F$ is insolation (in watts per square meter) and the Sun's insolation at Earth's orbit, $F_\oplus$, is 1361 W/m². Upper-atmosphere temperatures $T_{ua} > T_{eq}$ are essential for a Super-Earth atmosphere to flow out to space: this is possible because upper atmospheres efficiently absorb light at wavelength ≲ 100 nm, but do not readily reemit this light.

We adopt $s_{H2} = 2 \times 10^{-12}$ Pa$^{-1}$ (Hirschmann et al. 2012; Supplementary Information 1e), which for initial atmospheric pressure $P_{atm,init}$ = 50 kbar gives a total mass of H$_2$ (atmosphere plus dissolved-in-magma) of $7 \times 10^{23}$ kg (2% of planet mass). We neglect He, so our model primary atmosphere is slightly more soluble in magma and has a slightly lower molecular weight than in reality. We assume the crystal-melt partition coefficients are $D_{H2}$ = 0 (for simplicity) and $D_s$ = 0.02 (Supplementary Information 1f). Sub-Neptunes have a global shell of magma that freezes during conversion to a Super-Earth (Fig. S1). Planet thermal structure (below the photosphere, which is isothermal by assumption) is as follows (Fig. S2). The temperature at the top of the magma ocean, $T_{mai}$, is given by the atmosphere adiabat

$$(T_{mai}/T_{RCB}) = (P_{atm}/P_{RCB})^{(\gamma-1)/\gamma} \qquad (5)$$

where $T_{RCB}$ is the temperature at the radiative-convective boundary within the atmosphere (RCB), $P_{RCB}$ is the pressure at the RCB, and $\gamma$ is the adiabatic index. We assume $T_{RCB}/T_{eq}$ = 1, that the temperature jump at the magma-atmosphere interface is small, and that the magma ocean is isentropic. (If $T_{RCB}/T_{eq}$ = 1.5 in a 1000 K orbit then the planet cooling timescale is only < 1 Kyr). This basic model absorbs the planet cooling rate and the radiative opacity of the atmosphere into a single parameter, $P_{RCB}$ (Supplementary Information 1c); for more sophisticated models, see e.g. Marcq et al. (2017). As $P_{atm}$ decreases, $T_{mai}$ cools, and the magma crystallizes. Crystallization starts at great depth and the crystallization front sweeps slowly upward (e.g. Elkins-Tanton 2008). Volatiles enriched near the crystallization front due to the low solubility of volatiles in crystals will be stirred by fast magma currents (speed up to 10 m s$^{-1}$, Solomatov 2015) to the near-surface, where they form bubbles that pop and add gas to the atmosphere. Stirring and bubbling can degas the mantle down to at least ~100 GPa depths (e.g. Caracas et al. 2019; Solomotov 2015, and references therein). A smaller portion of $s$ will go into the crystals. This portion is shielded within the rock and available for later volcanic outgassing. We do not include liquid volatiles (e.g. clouds) or fluid-fluid immiscibility. We emphasize results for $T_{eq}$ < 1150 K, cold enough for silicates to condense (Woitke et al. 2018). The model is intended for super-Earths too hot for life.

Volcanic outgassing after the magma ocean has completely crystallized is guided by the results of Kite et al. (2009) (Fig. S3) (Supplementary Information 1h).

We approximate diffusive separation of $H_2$ and heavy gases as zero during sub-Neptune to super-Earth conversion. During conversion, escape proceeds too quickly for the *s* to settle out (Hu et al. 2015) (Supplementary Information 1g). On worlds that are cool enough for life, diffusive separation is important (Supplementary Information 1g). On worlds that are cool enough for life, diffusive separation can allow high-$\mu$ species to be retained in the atmosphere while H escapes, including in cases where *s* is a reducing species that is dissociated into easily-escaping H plus a heavy atom (e.g. $H_2O \rightarrow 2H + O$; Tian 2015). The secondary atmospheres that may be exsolved as the magma ocean crystallizes form more quickly (Elkins-Tanton 2011) than volcanically-outgassed atmospheres. Volcanically-outgassed atmospheres are produced on solid-state mantle homogenization timescales ($\tau >$ 1 Gyr; e.g., Gonnerman & Mukhopadhyay 2009, Saji et al. 2018). Because volcanic degassing of terrestrial planets is so slow/inefficient (e.g. Kaula 1999), we consider magma-ocean exsolution separately from volcanic outgassing. Specifically, we set rerelease of *s* from the solid mantle to zero so long as the exsolved atmosphere is present. With these approximations, for a given insolation *L*, planet mass $M_{pl}$, initial dose of *s*, and $P_{atm}$ at which that dose is applied, the output depends on how much atmosphere has been removed, but not on the speed (or process) of removal. In other words, the equations are time-independent.

**The small planet evolution sequence.**

We first model atmospheric evolution during magma ocean crystallization (Figs. 3-4). We show results for 6 $M_\oplus$ ($\approx$1.6 $R_\oplus$), corresponding to the largest (and therefore highest signal/noise) planets that commonly have densities consistent with loss of all $H_2$ (Rogers 2015). We use a total mass of high-$\mu$ volatile ($M_s$) = 3 × $10^{21}$ kg. This corresponds to the near-surface C inventory of Earth and Venus, scaled up to 6 $M_\oplus$ (and it is 2× the mass of Earth's ocean).

We drive the model by decreasing $P_{atm}$. In Figure 3, as $P_{H2}$ (blue dashed lines) falls, the magma ocean crystallizes, so that remaining volatiles go into the atmosphere (green), go into the solid mantle (maroon), or escape to space (black). We find that the main controls on the transition on exoplanets from primary to secondary atmospheres are *F* and the solubility ($s_s$) of the high-$\mu$ atmosphere constituent.

Fig. 3a shows a case with low $s_s$ ($10^{-11}$ Pa$^{-1}$), for a planet close to its host star ($F/F_\oplus$ = 240; $T_{eq}$ = 1100 K; typical for *Kepler* super-Earths). The magma ocean stays fully liquid until the atmospheric mass is reduced by 90%. (Release of dissolved-in-magma $H_2$ by bubbling is a negative feedback on atmospheric loss.) Crystallization begins at $P_{atm}$ = 2 kbar, and completes at ~100 bars. *s* is passively entrained (either as atoms or molecules) in the escaping $H_2$. We stop the run at $P_{atm}$ = 1 bar, so we do not track the removal of the last bit of the exsolved atmosphere. In this limit of small $s_s$, only 6 × $10^{16}$ kg × ($s_s$ / $10^{-11}$ Pa$^{-1}$) × ( $D_s$ / 0.02 ) is shielded within the solid mantle. Most of the *s* is lost to space before crystallization begins. The outcome is a bare rock with a volatile-starved solid mantle, incapable of much volcanic outgassing.

Fig. 3b shows results for the same $s_s$ as in Fig. 3a and a cooler orbit ($F/F_\oplus = 3$, $T_{eq} = 370$ K, intermediate between Mercury and Venus in our Solar System). In the cooler orbit, some crystals are present initially. With no fully-liquid stage, crystallization can start to shield $s$ within crystals as soon as atmospheric loss starts. The $s$ available for later volcanic outgassing in the cooler orbit case is 50 times greater.

Raising $s_s$ to $10^{-9}$ Pa$^{-1}$ (equivalent to 1 wt% solubility for 100 bars of partial pressure) favors secondary atmosphere occurrence (Figs. 3c-3d). More $s$ is dissolved in the magma, and so more $s$ partitions (during crystallization) into the rock, where it is shielded. Because $s$ is much more soluble in the magma than H$_2$, very little $s$ is initially in the atmosphere. Therefore, relatively little $s$ is carried away to space during H$_2$ removal and so more $s$ is available for exsolution during the final crystallization of the magma ocean. This can create a high-$\mu_{avg}$ atmosphere, which is easier to retain. Protection by differential solubility is enhanced by the hot orbit ($L/L_\oplus = 240$; Fig. 3c) as opposed to the cool orbit ($L/L_\oplus = 3$; Fig. 3d), because the magma ocean is long-lived for this case and so the volatiles are safely dissolved for longer. As a result, in the hot orbit case, enough high-$\mu$ species remain at final magma ocean crystallization to create an atmosphere with high-$\mu_{avg}$ (the green line crosses the blue line in Fig. 3c). Such a high-$\mu_{avg}$ atmosphere is easier to retain. The small amount of H$_2$ that remains in the atmosphere at this point can be lost by diffusion-limited escape.

Together, $F$ and $s_s$ have strong effects on the chance of secondary atmosphere occurrence. Volatiles that are shielded within rock are available for late volcanic outgassing. For high $s_s$ the mass of shielded volatiles tends toward the product of the initial inventory of $s$ and the solid-liquid partition coefficient $D_s$ (Fig. 4a). This is because almost all of the $s$ is in the magma until the magma ocean has almost completely crystallized. The effect of $F/F_\oplus$ on the mass of shielded volatiles is relatively modest for $s_s > 3 \times 10^{-11}$ Pa$^{-1}$. For $F/F_\oplus \leq 25$ and low $s_s < (\mu_s/\mu_{H2})s_{H2}$, the amount of the high-molecular-weight species that is shielded within rock is proportional to $s_s$. For low $s_s$, the mass of shielded volatiles decreases rapidly for hot orbits. In hot orbits the H wind can carry away $s$ for a long time before the magma ocean cools enough for solidification (and shielding).

When $\mu_{avg}$ is high, loss to space is less likely (Eqn. 1 and surrounding discussion). Fig. 4b shows how atmospheric $\mu_{avg}$ (after crystallization of the magma ocean is complete) depends on $F$ and $s_s$. For $s_s > 10^{-11}$ Pa$^{-1}$, a greater fraction of the $s_s$ is stored in the magma than is H$_2$. As a result, the atmosphere becomes more $s$-rich as the magma ocean crystallizes. The enrichment is especially strong for hot orbits, because for hot orbits there is more $s$ available to be exsolved: less $s$ has escaped to space before crystallization completes. For $s_s < 10^{-11}$ Pa$^{-1}$, most of the $s$ is stored in the atmosphere, and so the atmospheric $s$ mixing ratio is near-constant during atmosphere loss. (At $F/F_\oplus < 3$, diffusive separation favors high-$\mu_{avg}$ atmospheres; Supplementary Information 1 g).

Cooler orbits favor volcanic outgassing but hotter orbits permit exsolved high-molecular-weight atmospheres. To determine which effect is more important for the chance of seeing a secondary atmosphere on a super-Earth, we turn to a time-dependent model.

**Where are planets today on the small planet evolution sequence?**

We map planet evolution onto time and host-star mass (Fig. 5). The atmosphere loss that converts sub-Neptunes into super-Earths could be driven by photoevaporation, impact erosion, or accretion energy (e.g. Owen & Wu 2017, Gupta & Schlichting 2019, Kegerreis et al. 2020). Here we consider photoevaporation due to X-ray and Extreme Ultraviolet flux (XUV). $F_{XUV}$ plateaus at $\sim 10^{-3} \times$ total $F$ for planets around young stars, switching to a power-law decay at < 0.1 Gyr for Solar-mass stars ($F_{XUV} = 3 \times 10^{-6}\ F$ at Earth today) (Supplementary Information 1a). The plateau of high $F_{XUV}/F$ is longer at red dwarf (M) stars (≥0.3 Gyr long for ≤ 0.5 $M_\odot$). Therefore, we expect that (for a given $T_{eq}$) planets orbiting M-stars will have lost more atmosphere (e.g. Lammer et al. 2007).

$F_{XUV}$ drives atmospheric loss (rate $dM_{atm}/dt$) (Supplementary Information 1b). Nebula-composition atmospheres do not cool efficiently, leading to high $T_{ua}$ that favors hydrodynamic escape (Murray-Clay et al. 2009). For nebular-composition atmospheres, we use the following:

$$dM_{atm}/dt = \varepsilon\ R_{pl}\ (R+z(t))^2\ F_{XUV} / (G\ M_{pl}) \qquad (6)$$

where $\varepsilon$ is an efficiency factor. For high-$\mu$ atmospheres we adopt the loss fluxes of a $CO_2$ atmosphere (Tian 2009) as an example of a strong coolant. The model of Tian (2009) and Tian et al. (2009) includes dissociation of $CO_2$ under high XUV levels, and the escaping material is atomic C and O. The model of Tian (2009) predicts low $\varepsilon$ for $CO_2$ atmospheres, and negligible hydrodynamic escape for $F_{XUV}$ < 0.6 W m$^{-2}$ (= 150× the value on Earth today) (Fig. S4). At intermediate compositions we interpolate using a logistic function (Supplementary Information 1b).

Fig. 5 shows atmosphere thickness vs. time for a Solar-mass star and $P_{atm,init}$ = 50 kbar. High-$\mu_{avg}$ atmospheres can only persist at a narrow range of $F$. (A similar pattern is seen for low-mass stars; Figure S6). Worlds far from the star receive a low XUV flux and stay as sub-Neptunes. Worlds in hotter orbits lose their atmospheres completely. They may undergo a rapid increase in $\mu_{avg}$ (blue to yellow in Fig. 5a), but the resulting high-$\mu_{avg}$ atmosphere is almost always short-lived. Why is this? According to the pure-$CO_2$ model of Tian (2009), $dM_{atm}/dt$ for high molecular weight atmospheres on super-Earths has a threshold at $\sim$150× $F_{XUV}/F_{XUV,\oplus}$, below which loss is much slower. However, in most cases the primary atmosphere is lost when $F_{XUV}$ is still very high so any exsolved secondary atmosphere is swiftly lost. For low $s_s$, the atmosphere has a composition that is always $H_2$-dominated (by number), so exsolved high-$\mu$ species are lost in the H wind (Fig. 5b).

We conclude that exsolved atmospheres are rare. This conclusion is robust because we use parameters that are favorable for an exsolved atmosphere. For example, we use a high solubility-in-magma for the high-$\mu$ species, but our high-$\mu$ escape-to-space parameterization is for a species ($CO_2$) whose solubility-in-magma is low.

**Revival of secondary atmospheres by volcanic outgassing.**

Volcanic outgassing can regenerate the atmosphere of a planet. We do not know whether or not this process actually occurs on rocky exoplanets. We use a basic model to explore this process. We use a time-dependent rate-of-volcanism model for super-Earths (Kite et al. 2009; Supplementary Information 1h) that is tuned to the rate of $CO_2$ release at Earth's mid-ocean ridges (12±2 bars/Gyr; Tucker et al. 2018). The rate of outgassing is adjusted downward to account for loss of volatiles during sub-Neptune-to-super-Earth conversion, assuming 50% of worlds have $s_s = 10^{-9}$ Pa$^{-1}$, and 50% of worlds have $s_s = 10^{-11}$ Pa$^{-1}$. We neglect atmosphere re-uptake to form minerals (e.g., Sleep & Zahnle 2001), because our focus is on worlds that are too hot for aqueous weathering. This omission is conservative relative to our conclusion that volcanically-outgassed atmospheres on hot rocky exoplanets are uncommon. Fig. 6 (gray curve) outlines the region within which, in our model, volcanic outgassing will build up a secondary atmosphere. The line of revival sweeps toward the star over time, because the rate of volcanic degassing falls off more slowly with time than does the star's XUV flux. Volcanic revival of the atmosphere is difficult for planets around $M = 0.3\ M_\odot$ stars, but easier for rocky planets around solar-mass stars. The results are sensitive to changes in $s_s$, and to the choice of XUV flux models (Figs. S9-S11, S17). Fig. S17 also shows results for volcanism at a constant Earth-scaled outgassing rate. Fig. 6 also shows the atmosphere presence/absence line for rocky worlds that start with no $H_2$, and with all high-molecular-weight volatiles in the atmosphere (red curves). Such "intrinsically rocky" worlds retain residual secondary atmospheres over a wider range of conditions than do worlds that start as sub-Neptunes. The line of atmosphere loss for these residual atmospheres sweeps further away from the star with time.

**Discussion.**

Our model results are sensitive to the rate of XUV-driven mass loss. The XUV flux of young Solar-mass stars varies between stars of similar age by a factor of 3-10 (e.g. Tu et al. 2015). Stars with low XUV flux are more likely to host planets with atmospheres (Fig. S11). Escape of $N_2$ or $H_2O$ is likely faster than the Tian (2009) loss rate estimate (which is for pure-$CO_2$ atmospheres) that is used in our model (Zahnle et al. 2019, Johnstone et al. 2018). Improved knowledge of escape rate will require more escape rate data and XUV flux data (e.g. Bourrier et al. 2018, Ardila et al. 2018). Currently, XUV flux data as a function of star mass and star age are limited, and upcoming space missions such as SPARCS will gather more data (Ardila et al. 2018, Linsky et al. 2019).

Our model considers only thermal loss. However, solar wind erosion can remove atmospheres that are already thin (Dong et al. 2018). A possible example is Mars over the last ~4 Ga. Planetary dynamos can under some circumstances suppress solar-wind erosion (e.g. Gunnell et al. 2018). Mars had a dynamo prior to 4 Gya, and Earth (but not Venus) has one today.

Solubility of gas in magma varies between species. Carbon-bearing volatile species are very insoluble in magma. $H_2O$'s solubility in basaltic magma at $H_2O$ partial pressure 0.03 GPa is ~2 wt%; linearizing, this gives solubility $s_s = 7 \times 10^{-10}$ Pa$^{-1}$ (e.g. Papale 1997). HCl is even more soluble in magma than $H_2O$ (Fegley 2020), so a Cl-rich initial composition would have a greater chance of forming an exsolved high-$\mu$ atmosphere, although we still predict it would be short-lived. The effects on atmosphere prevalence of 100-fold reduction in solubility are shown in Figs. S9-S11. Better constraints on solubility at $T > 2000$ K are desirable (e.g. Guillot & Sator 2011).

Volcanism on super-Earths should wane over gigayears, according to models. We need more data to test these models. In models, the rate of decrease of volcanism depends on whether or not the planet has plate tectonics, on planet mass, and on mantle composition (e.g. Stevenson 2003, Unterborn et al. 2015, Dorn et al. 2018, Byrne 2019; Supplementary Information). Two effects, of uncertain relative importance, are ignored in our volcanism model. The first is a buffering effect: if the mantle is volatile-rich, then magma is produced more easily (Médard & Grove 2006), but if the mantle is volatile-poor, then the melt rate is reduced (reducing the rate of volatile loss). The second is a potential fine-tuning issue: if volcanic degassing is very rapid, then volatiles will be released from the protective custody of the mantle before atmosphere loss process have lost their bite.

Overall, our choices of $M_{pl}$, $D_s$, and $s_s$ tend to favor the existence of volcanically outgassed atmospheres. Even with these choices atmospheres are usually not stable for planets at $T_{eq} > 500$ K around M-stars. So, our conclusions are broadly unfavorable for atmospheres on rocky exoplanets at $T_{eq} > 500$ K around M-stars. However, this conclusion is moderated by the possibility (discussed next) that worlds with abundant H$_2$O exist close to the star.

**Observational tests.**

Atmospheres on rocky exoplanets can now be detected (e.g. Demory et al. 2016, Kreidberg et al. 2019). Theory predicts that retaining an atmosphere should be harder on planets orbiting low-mass stars and the present study extends that prediction to super-Earths that form as sub-Neptunes. A test of this theory would have major implications for habitable zone planets. If this prediction fails, that would suggest that M-star rocky exoplanets formed more volatile-rich than rocky exoplanets orbiting Sun-like stars (Tian & Ida 2015).

It is possible that some planets form with more volatiles than can be removed by loss processes (Bitsch et al. 2019). Some models predict formation of hot Super-Earths with 1 wt % - 30 wt% H2O, either by accretion of volatile-rich objects (for example, extrasolar analogs to CI/CM chondrites), or by planet migration (e.g. Raymond et al. 2018, Bitsch et al. 2019). Such volatile-rich worlds are hard to distinguish from bare-rock planets using current data. XUV-driven loss can remove at most a few wt% of an $M = 6$ $M_\oplus$ planet's mass for $T_{eq} < 1000$ K. Our model implies that a planet in the "no atmosphere" zone of Fig. 6 with a JWST detection of H$_2$O-dominated atmosphere is more volatile-rich than Venus and Earth. Possible volatile-rich worlds include planets that have radii ~0.2 $R_\oplus$ greater than expected for Earth-composition (Fig S12) (Turbet et al. 2019).

Fig. 6 enables the following testable predictions. Since atmospheres close to the star can only persist if the initial H$_2$O inventory is high, N$_2$/NH$_3$ should be diluted to very low mixing ratio for such atmospheres. If a super-Earth-sized planet has an atmosphere, then planets at greater semimajor axis in the same system should also have an atmosphere. Starting out as a sub-Neptune is unfavorable for atmosphere persistence, so systems where the planets formed intrinsically rocky should have statistically more atmospheres. Multiplanet systems enable strong tests because uncertainty in time-integrated stellar flux cancels out.

**Conclusions.**

A large fraction of rocky exoplanets on close-in orbits (closer to their star than the Habitable Zone) were born with thick (>10 kbar) $H_2$-dominated (primary) atmospheres but have since lost their $H_2$. In our model these $T_{eq}$ > 400 K exoplanets almost never transition smoothly to worlds with high-molecular weight atmospheres (Fig. 7). Instead, the high-molecular-weight species are usually carried away to space by the H wind. Volcanic outgassing is an alternative source for a high-molecular-weight atmosphere. Revival of a bare-rock planet by volcanic outgassing gets easier with time, because atmosphere loss slows down rapidly with time, but atmosphere supply by volcanism decays slowly. However, volcanic outgassing is also enfeebled by early loss of biocritical volatiles via the H wind. Overall, for a given initial dose of high-molecular-weight species, atmospheres are less likely on hot rocky exoplanets that were born with thick $H_2$-dominated atmospheres. Many uncertainties remain, the most important of which is the initial planet volatile content. Within our model, for planets that orbit Solar-mass stars, super-Earth atmospheres are possible at insolations much higher than for planet Mercury. For planets that orbit ~0.3 $M_\odot$ stars, secondary atmospheres at much higher insolation than planet Mercury in our solar system are unlikely unless the planet formed $H_2$-poor, or includes a major (~1 wt%) contribution of solids from beyond the water ice line (water world).

**Data availability:** All of the code for this paper, together with instructions to reproduce each of the figures and supplementary figures, can be obtained via the Open Science Framework at https://osf.io/t9h68 or by emailing the lead author.

**Acknowledgements:** We thank two reviewers for accurate and useful reviews. We thank B. Fegley, Jr., L. Schaefer, L. Rogers, E. Ford, and J. Bean (discussions). Grants: NASA Exoplanets Research Program (NNX16AB44G).

**Figures.**

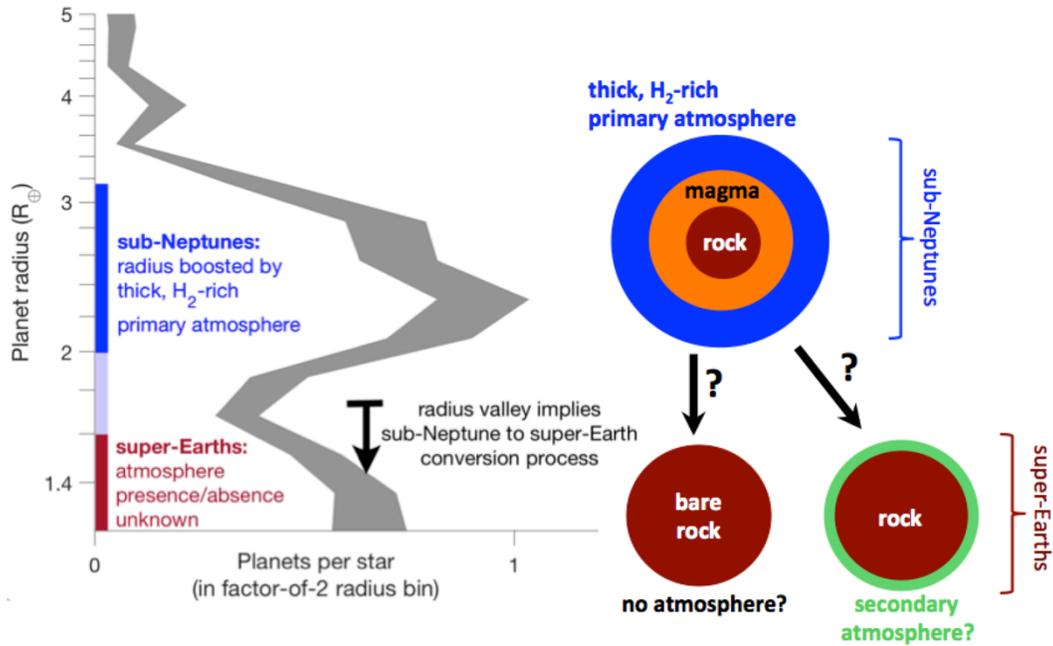

**Figure 1.** The exoplanet abundance histogram (gray band, for orbital periods < 100 days, corrected for detection biases; Fulton & Petigura 2018). Two classes of small exoplanet are seen: volatile-rich sub-Neptunes and rocky super-Earth-sized exoplanets. Sub-Neptune to super-Earth conversion is implied by the data and may be the way that most super-Earths form.

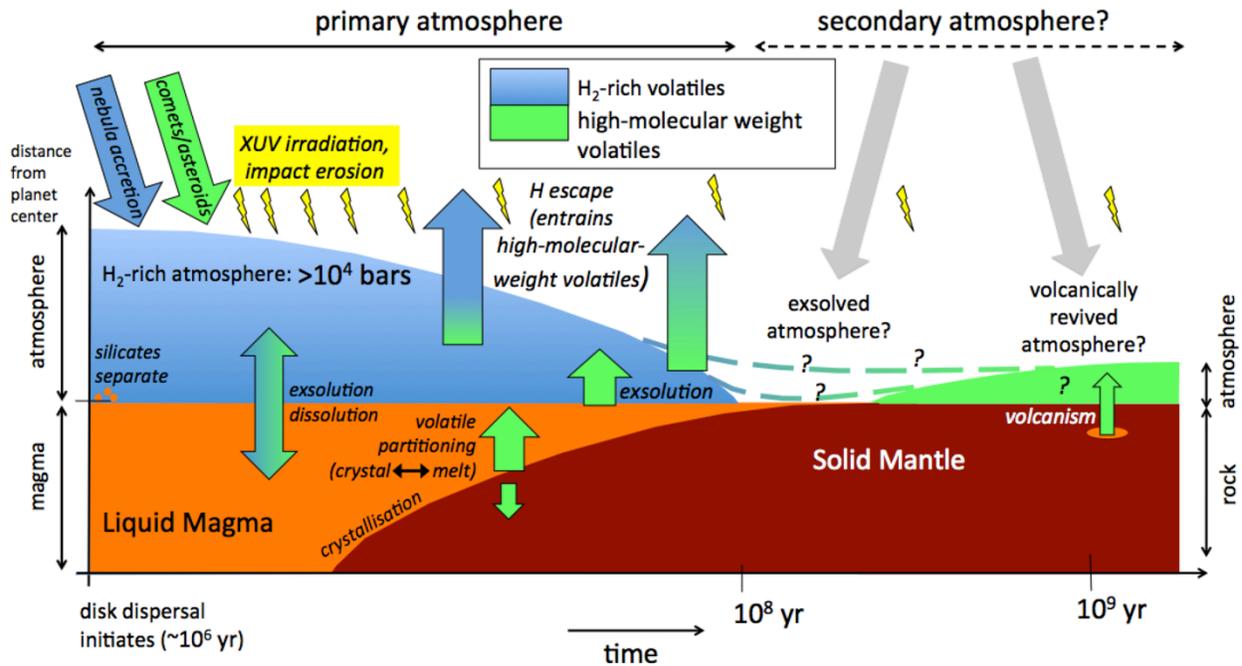

**Figure 2.** Processes (italics) and reservoirs (upright font) in our model. Atmosphere-interior exchange is central to the transition from primary to secondary atmospheres. Timescales are approximate.

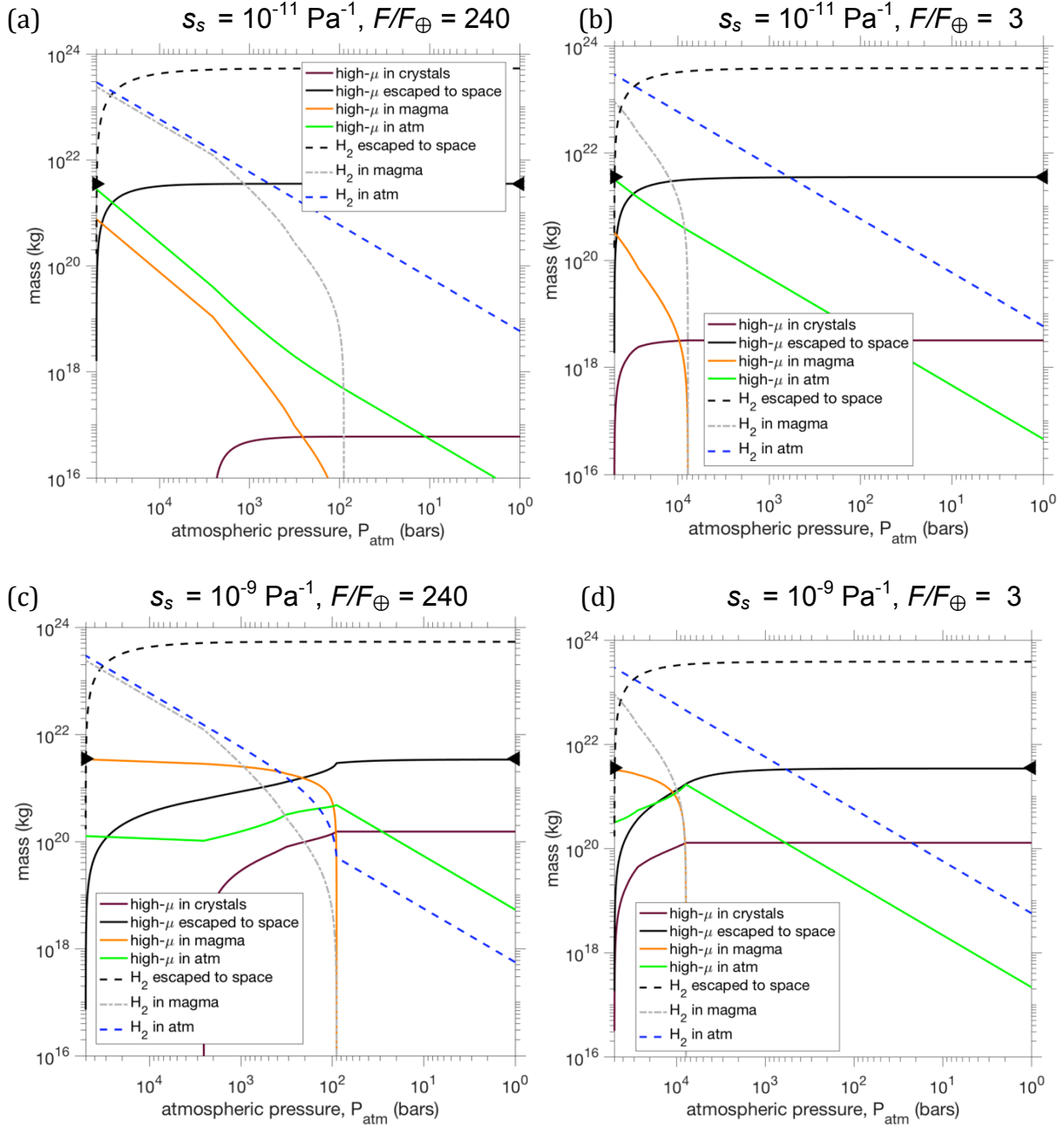

**Figure 3.** The small planet evolution sequence, for 6 $M_\oplus$. Black triangles correspond to total initial inventory of high-$\mu$ species. (a) Solubility of high-$\mu$ species in magma ($s_s$) = $10^{-11}$ Pa$^{-1}$, insolation normalized to Earth's insolation ($F/F_\oplus$) = 240 (planet equilibrium temperature 1095 K). (b) $s_s$ = $10^{-11}$ Pa$^{-1}$, $F/F_\oplus$ = 3 (planet equilibrium temperature = 360 K). (c) $s_s$ = $10^{-9}$ Pa$^{-1}$, $F/F_\oplus$ = 240. (d) $s_s$ = $10^{-9}$ Pa$^{-1}$, $F/F_\oplus$ = 3.

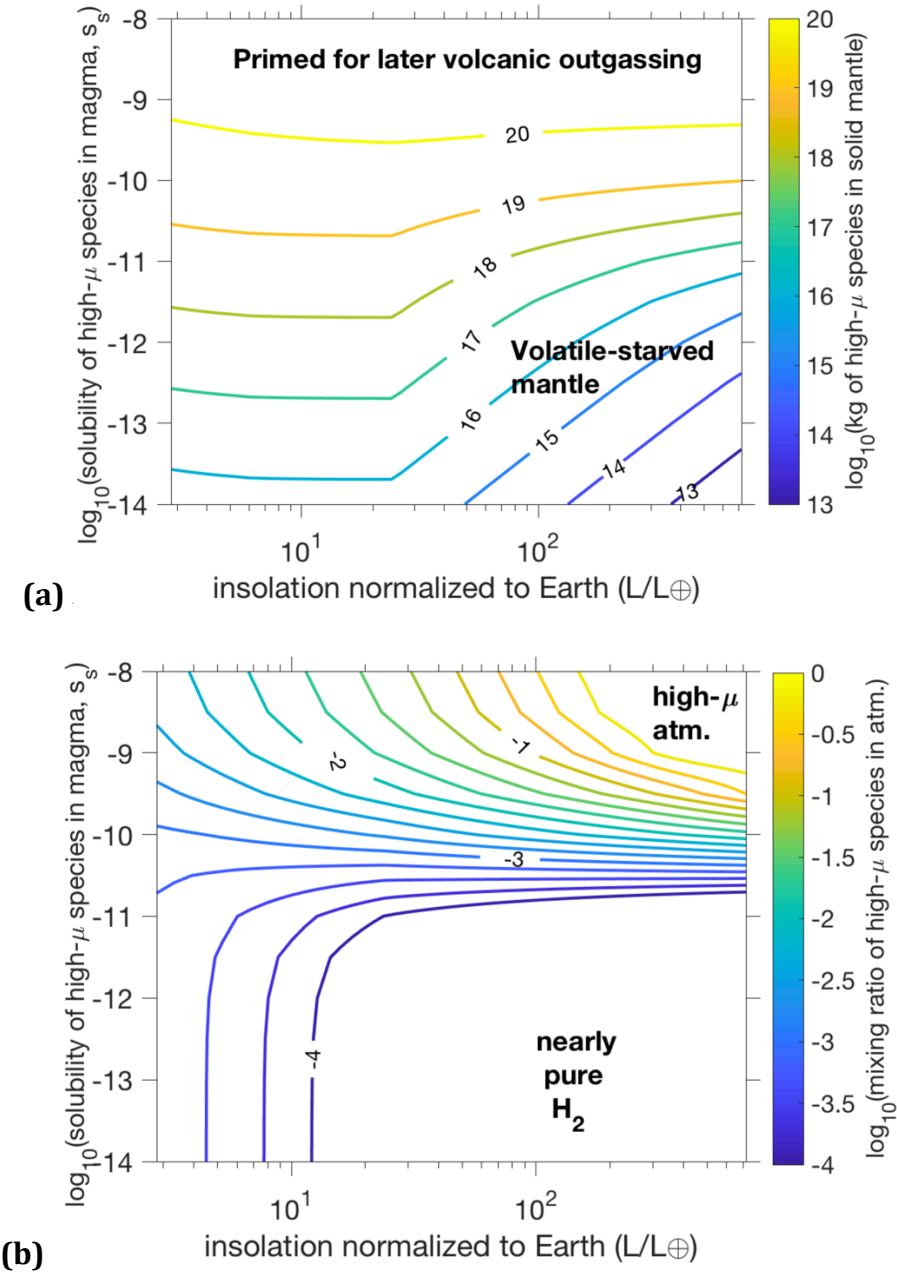

**Fig. 4. (a)** How the mass of volatiles shielded within the solid mantle depends on *F* and $s_s$. **(b)** How atmospheric mean molecular weight (after crystallization of the magma ocean is complete) depends on *F* and $s_s$.

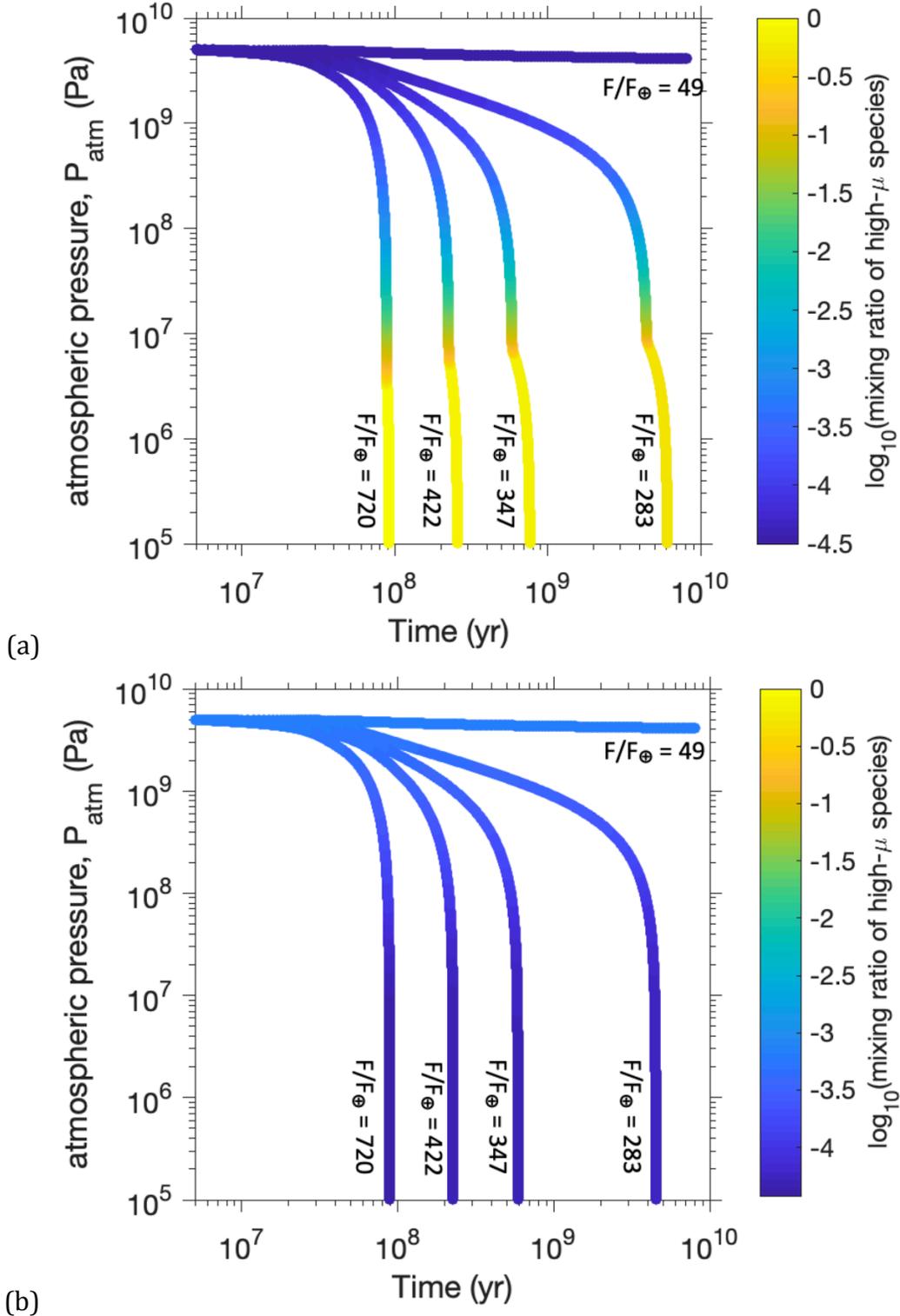

**Figure 5.** Time-dependent results for planets orbiting a Solar-mass star. (Fig. S6 shows the results for a 0.3 $M_\odot$ star.) **(a)** Atmospheric pressure vs. time for $s_s = 10^{-9}$ Pa$^{-1}$, (from top to bottom) $F/F_\oplus = \{49, 283, 347, 422, 720\}$, corresponding to planet equilibrium temperature ($T_{eq}$) = $\{735, 1140, 1200, 1275, 1440\}$ K. **(b)** As (a), but for $s_s = 10^{-11}$ Pa$^{-1}$.

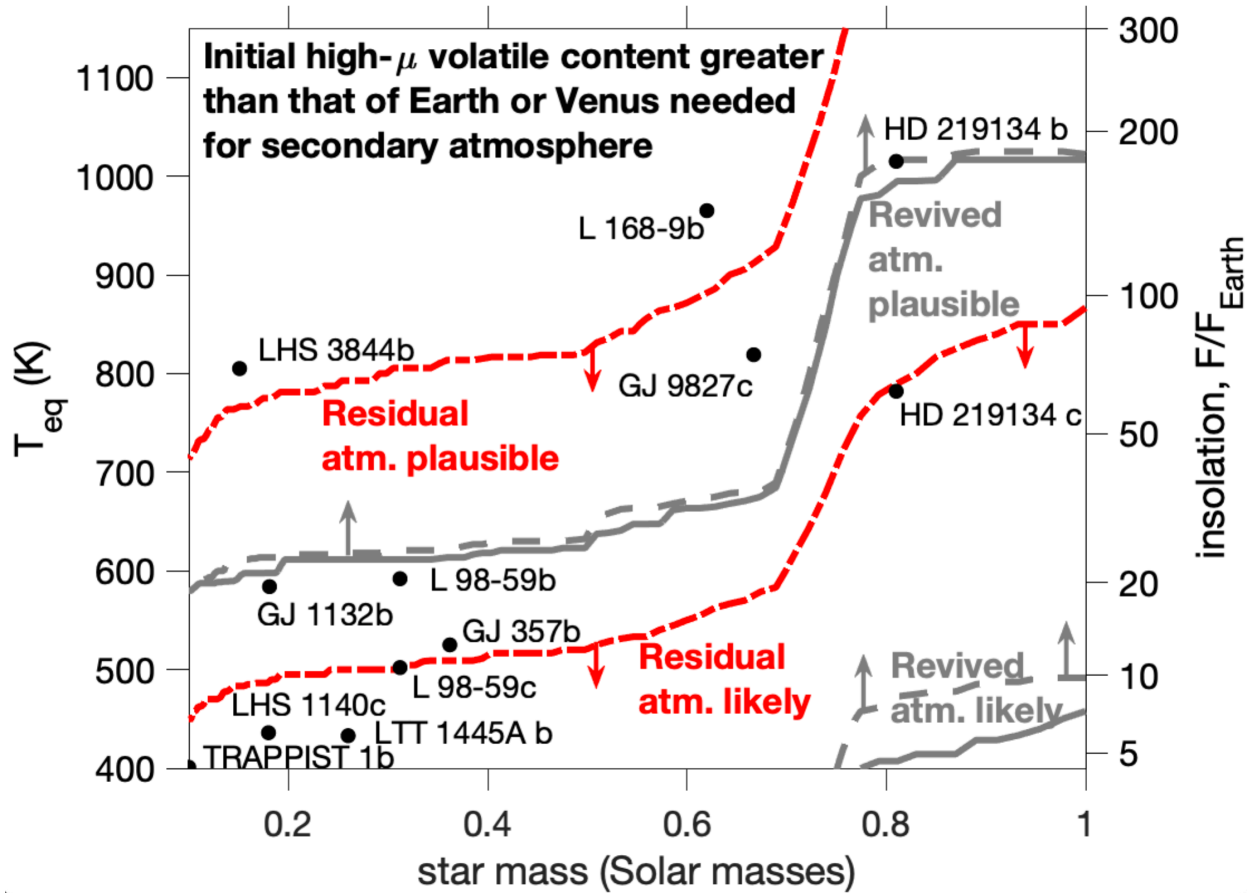

**Fig. 6.** Secondary atmosphere presence/absence model output for 6 $M_\oplus$ (higher planet mass favors atmosphere retention). The red lines and gray lines show atmosphere presence/absence contours for two different scenarios. The red lines show atmosphere retention thresholds after 3.0 Gyr for the case where all volatiles are in the atmosphere initially and there is no primary atmosphere; the 16th and 84th percentiles are shown, for varying XUV flux (by ±0.4 dex, 1σ; Loyd et al. 2019) relative to the baseline model following the results of Jackson et al. (2012) and Guinan et al. (2016) (Supplementary Information 1a). The red lines move away from the star over time (red arrows). The gray lines show the 16th and 84th percentiles for exhibiting an atmosphere after 3.0 Gyr for the case where volcanic outgassing rebuilds the atmosphere from a bare-rock state. The solid gray lines are for stagnant-lid tectonics and the dashed gray lines are for plate tectonics. The lines of atmospheric revival sweep towards the star over time (gray arrows) because the rate of volcanic degassing falls off more slowly with time than does the star's XUV flux. In each case the atmosphere/no-atmosphere threshold is 1 bar. The black symbols show known planets that may be tested for atmospheres using JWST (Koll et al. 2019). For any individual planet, star-specific XUV-flux estimates, star age, and the planet's mass, should be combined to make a more accurate estimate than is possible using this overview diagram. Figs. S8 and S16-17 show further details.

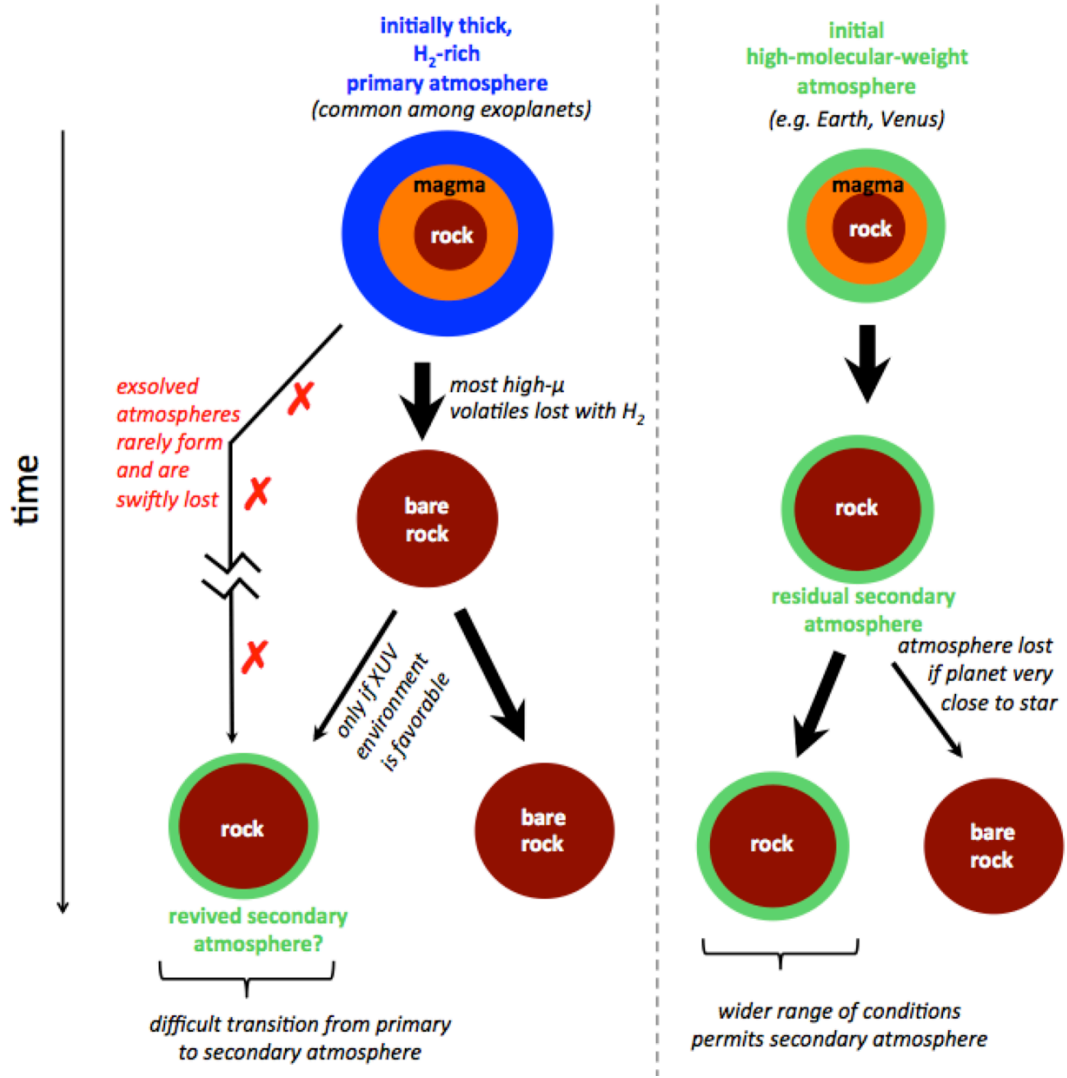

**Fig. 7.** Graphical summary. The left column corresponds to atmosphere revival (gray lines in Fig. 6) and the right column corresponds to residual atmospheres (red lines in Fig. 6). For worlds in $T_{eq} > 400$ K orbits that start as sub-Neptunes, formation of a high-$\mu$ atmosphere is unlikely, unless the planet starts with abundant high-$\mu$ volatiles.

**References.**

**SI Appendix.**

*1. Model description.*

Atmosphere evolution involves many processes (Catling & Kasting 2017). This is (to our knowledge) the first paper on the transition on exoplanets from primary to secondary atmospheres. Therefore our approach is to keep the treatment of each process simple.

We consider star ages from 5 Myr, which is a typical time for nebular disk dispersal, up to 8 Gyr. We consider star masses from 0.1-1 $M_\odot$. We frequently state results for 0.3 $M_\odot$, because these are expected to be the most common host stars for worlds detected by the TESS mission (e.g. Sullivan et al. 2015, Huang et al. 2018). We emphasize planet mass 6 $M_\oplus$ (bare-rock radius ≈ 1.6 $R_\oplus$), because these are the largest (and therefore highest signal/noise) worlds that commonly have densities consistent with loss of all $H_2$ (Rogers 2015). The silicate mass fraction is held fixed at 2/3, with the balance consisting of the Fe/Ni metal core. This silicate mass fraction is based on Solar System data. Our results are only weakly sensitive to reasonable variations in the silicate mass fraction. The total mass of $H_2$ is small compared to planet mass (Lopez & Fortney 2014). The non-$H_2$ volatile mass fraction is also assumed to be small, consistent with data (e.g. van Eylen et al. 2018).

*1a. Drivers of atmosphere loss.*

Candidate drivers for the atmosphere loss that converts sub-Neptunes into super-Earths are photoevaporation, impact erosion, and core luminosity (e.g. Inamdar & Schlichting 2016, Zahnle & Catling 2017, Burger et al. 2018, Ginzburg et al. 2018, Owen & Wu 2017, Biersteker & Schlichting 2019, Kegerreis et al. 2020, Denman et al. 2020). We focus on photoevaporation in this study. Photoevaporation has been directly observed for sub-Neptune exoplanets (Bourrier et al. 2018), fits almost all of the data (e.g. Van Eylen et al. 2018), and is relatively well-understood (e.g. Murray-Clay et al. 2009). Other mechanisms are important at least for a small number of planetary systems (e.g. Owen & Estrada 2020, Loyd et al. 2020).

The rate of photoevaporation is set by the X-ray / Extreme Ultraviolet (XUV) flux, $F_{XUV}$ (e.g. Murray-Clay et al. 2009, but see also Howe et al. 2020). XUV is generated by hot regions near the star's surface. These regions are heated by magnetic fields whose importance declines as the star sheds angular momentum, through the stellar wind, over time. Thus the star's XUV luminosity $F_{XUV}$ declines over time. Early in a star's history, the star's XUV luminosity $L_{XUV}$ reaches a plateau value (referred to as "saturation") of between $10^{-4}\times$ and $10^{-3}\times$ the star's total, bolometric luminosity (*L)*.

Estimating $L_{XUV}$ involves combining direct measurements for stars of similar spectral type and age, and interpolation over wavelength regions where star UV is absorbed by the interstellar medium. Data on $L_{XUV}$ is synthesized by Linsky (2019). Lopez & Rice (2018) and Luger & Barnes (2015) propose power-law decays for Sun-like stars. Tu et al. (2015) quantify the variability of X-ray emission for stars of Solar mass. Guinan et al. (2016) compile $L_{XUV}$ data for low-mass stars.

We used two approaches to interpolating in $L_{XUV}/L$ as a function of star mass and time. In one approach, we used Figure 5 from Selsis et al. (2007) for the X-ray flux, obtaining the XUV flux from the X-ray flux using the top row of Table 4 in King et al. (2018). Selsis et al. (2007) assume a constant $L_{XUV}/L$ of $10^{-3.2}$ during saturation. Selsis et al. (2007) do not attempt to quantify the variation of X-ray flux between stars of the same age and the same stellar mass. In the other approach, we used Jackson et al.'s (2012) fits for $L_{XUV}/L$ for $M \geq 0.5\ M_\odot$, patching to Guinan et al.'s (2016) fits for $M < 0.5\ M_\odot$. We assumed that Guinan et al.'s (2016) fits applied for $M = 0.35\ M_\odot$ and extrapolated to other M-stars. We used standard models for $L$ and star radius (Baraffe et al. 2015) (Fig. S13). The $L_{XUV}/L$ output from the two approaches is shown in Fig. S14. The zones of secondary atmosphere loss and revival from these approaches are shown in Fig. S9-S11. For both approaches, we also calculated how results would vary between stars of different $L_{XUV}/L$, assuming a 1-standard-error scatter in $L_{XUV}/L$ of ±(0.4-0.5) dex (following Loyd et al. 2020) (Fig. 6, S16). We use an upper wavelength cutoff of 91.2 nm (13.6 eV), corresponding to the lowest-energy photon that can ionize H. However, longer-wavelength light can still contribute to atmospheric escape, especially as $\mu_{avg}$ rises. Future work using more sophisticated escape models (e.g. Wang & Dai 2018) might investigate the effect of switching on or off various UV spectral bands at lower energies than 13.6 eV.

*1b. Atmosphere loss rates.*

For pure-$H_2$ atmospheres, we set

$$dM_{atm}/dt = \varepsilon\, R_{pl}\, (R+z)^2\, F_{XUV} / (G\, M_{pl}) \qquad (S1)$$

(Eqn. 6 from main text) where the heating efficiency factor, $\varepsilon = 0.15$. $\varepsilon = 0.15$ is a common choice in atmospheric evolution models. Models indicate that $\varepsilon = 0.1$-$0.2$ (e.g. Shematovich et al. 2014). As the planet's atmosphere loses mass and cools, it shrinks (Lopez & Fortney 2014). This acts as a negative feedback on loss in part because smaller planets intercept fewer XUV photons. To relate the atmosphere mass $M_{atm}$ to the planet radius ($R+z$), we used the transit-radius tables of Lopez & Fortney (2014). The homopause radius could be bigger than the transit radius by typically 5-25% (Malsky & Rogers 2020).

For our pure-high-molecular-mass atmosphere endmember loss rate, we use the loss rate in molecules/sec/cm$^2$ calculated for a pure-$CO_2$ atmosphere on a super-Earth by Tian (2009). This is an endmember of a high-$\mu_{avg}$ atmosphere that can self-cool efficiently. This loss rate is multiplied by $\mu_{avg}$, and by the area of the surface-atmosphere interface, to get the loss rate in kg/s. Our intent is to bracket the likely down-shift in atmospheric loss as the atmosphere evolves to high-$\mu_{avg}$. We extrapolate beyond the XUV-flux limits shown in Figure 6 of Tian (2009) linearly in loss rate for very large XUV fluxes, and linearly in the log of loss rate for very small XUV fluxes. High-$\mu_{avg}$ atmospheres of different composition may shed mass faster (or slower) than the pure-$CO_2$ case (Johnstone 2020, Johnstone et al. 2018, Zahnle et al. 2019).

For atmospheres of intermediate composition, we interpolate between the energy-limited formula ($dM_{atm}/dt \propto F_{XUV}$) to the exponential cutoff in $F_{XUV}$ for high-$\mu_{avg}$ atmospheres ($dM_{atm}/dt \approx 0$ below a critical $F_{XUV}$) found by Tian (2009). Constraints for hydrodynamic loss rates of atmospheres of intermediate composition are limited (Kulikov et al. 2007, Johnstone et al. 2018). In the absence of better constraints, we use a logistic curve in log (escape rate) and log (mixing ratio) :

$$X_{s,atm} = (\mu_{avg} - \mu_p) / (\mu_s - \mu_p); \tag{S2}$$

$$Y = 1 / (1 + \exp(-k_1 \log_{10} X_{s,atm} - \log_{10} k_2 )) \tag{S3}$$

with $k_1$ = 8 and $k_2$ = 0.2. We think that these parameter choices understate the $X_{CO2}$ needed to suppress escape (i.e. favor the survival of high-$\mu$ atmospheres), which is conservative in terms of our conclusion that atmosphere survival is unlikely. Next we set

$$Z = \log_{10}(dM_{atm}/dt)_p - Y \log_{10} (dM_{atm}/dt)_p - \log_{10} (dM_{atm}/dt)_s \tag{S4}$$

$$(dM_{atm}/dt)_{mixed} = 10^Z \tag{S5}$$

where $(dM_{atm}/dt)_p$ is the energy-limited escape rate with $\varepsilon$ = 0.15, $(dM_{atm}/dt)_s$ is from Tian (2009), and $(dM_{atm}/dt)_{mixed}$ is the loss rate for impure atmospheres.

### 1c. Planet thermal structure.

Planet thermal structure in our model is defined by the temperature at the atmosphere's radiative-convective boundary ($T_{RCB}$) and the temperature at the magma-atmosphere interface, $T_{mai}$. $T_{RCB}$ and $T_{mai}$ are related by the thickness of the convecting (adiabatic) envelope, which (in pressure units) is ($P_{atm} - P_{RCB}$), and by the adiabatic index $\gamma$ (held constant at 4/3). We do not consider changes in $\gamma$ with changes in atmospheric composition.

Planets form hot and cool over time. Cooling involves energy transport through the atmosphere by radiation and/or convection. If, at a given optically thick level of the atmosphere, all cooling is accomplished by radiation, then the radiative temperature gradient is

$$dT/dr = - (3 \kappa \rho / 15 \sigma T^3) (L_{int} / 4 \pi r^2) \tag{S6}$$

(e.g. Rogers et al. 2011), where $\kappa$ is the local Rosseland-mean opacity, $\rho$ is the local atmospheric density, $T$ is local temperature, $\sigma$ is the Stefan constant, and ($L_{int} / 4 \pi r^2$) is the internal luminosity corresponding to the cooling of the planet (where $r$ is the distance from the center of the planet). If the radiative $dT/dr$ from Eqn. (S6) exceeds the convective adiabatic temperature gradient, then convection takes over and $dT/dr$ is reduced to a value slightly greater than the adiabatic lapse rate. Thus, as the planet cools, and $L_{int}$ decreases, the radiative zone expands ($P_{RCB}$ moves to greater depths) (e.g. Bodenheimer et al. 2018,

Vazan et al. 2018). We set $P_{RCB}$ = 10 bars and neglect $P_{RCB}$ change with time, which means our magma oceans are slightly longer-lived than they would be in reality. On the other hand, we use the 1 Gyr $T_{eq}$ value throughout the calculation, when for cool stars $T_{eq}$ will be higher earlier in the planet's life. This simplification tends to shorten model magma ocean lifetime relative to reality.

We assume full redistribution of energy absorbed on the dayside to the entire (4π) area of the planet. Our model planets are also isothermal with altitude above the radiative-convective boundary. In effect, we neglect the temperature difference between the planet photosphere (level at which the optical depth is approximately unity) and the radiative-convective boundary. At the top of the atmosphere we assume zero albedo. This is reasonable for cloud-free atmospheres of planets orbiting cool stars, although for Solar-mass stars the albedo will be ~0.2, considering only $CO_{2(g)}$ and $H_2O_{(g)}$ opacity (Pluriel et al. 2019).

Our focus is interior to the habitable zone and for thick atmospheres. For such worlds atmospheric retention would result in magma oceans that persist for Gyr (Hamano et al. 2013, Vazan et al. 2018, Fig. S1). Thus, interior to the habitable zone, atmosphere loss is the rate-limiter for magma ocean crystallization (Hamano et al. 2013, Hamano et al. 2015, Lebrun et al. 2013). In principle our model could lead to an equilibrium situation where a few-hundred-bar high-$\mu$ atmosphere sustains a magma ocean indefinitely. However such outcomes in our model (which assumes all volatiles are delivered early) require fine-tuning.

During sub-Neptune-to-super-Earth conversion, planet cooling is driven by atmospheric removal. As the overburden pressure is removed, the remaining atmosphere cools adiabatically. This causes a temperature gradient between the magma and the atmosphere, and heat flow from the magma into the atmosphere soon warms the atmosphere because the atmosphere heat capacity is small compared to the magma ocean heat capacity. This raises $T_{RCB}$, and because $T_{RCB}/T_{eq}$ > 1 cools the planet quickly, this cools the planet. All of these processes occur continuously in our simplified model: we assume the magma temperature tracks the loss of atmosphere. In effect we assume that convection swiftly reduces super-adiabatic temperature gradients and that the boundary layer temperature contrast at the top of the liquid magma (viscosity < $10^{-1}$ Pa s) is small. More detailed models of sub-Neptune thermal evolution include Rogers et al. (2011), Howe & Burrows (2015), Vazan et al. (2018), and Bodenheimer et al. (2018).

For cooler orbits in our model, some crystals are present initially. This incomplete-magma-ocean initial condition is appropriate if the planet is assembled by impacts of objects that are Mars-sized or smaller (e.g. planetesimals, pebbles, or planetary embryos), and the planet cools down to the adiabat between impacts.

### *1d. Magma ocean crystallization.*

Magma ocean crystallization shields dissolved volatiles from loss. The magma ocean crystallizes (losing mass to crystals) as the planet cools. The magma ocean mass at a given $T_{mai}$ depends on conditions deep within the silicate layer (higher $T$ and much higher $P$ than at the magma-atmosphere interface). The silicate solidus ($T$ vs. $P$ for 0% melt fraction) and liquidus ($T$ vs. $P$ for 100% melt fraction) are curved in $T$-$P$ space. As a result, for a steady decline in $T_{mai}$, the magma mass will decrease first quickly and then slowly. To track this effect, we used the solidus, the liquidus, and the adiabat of Andrault et al. (2011) (Fig. S15). We assume that the melt fraction increases linearly with increasing temperature between the solidus and the liquidus, and neglect chemical fractionation during crystallization. We also neglect the reduction in the solidus $T$ due to volatile enrichment as crystallization proceeds. More detailed models of magma ocean crystallization include Nikolau et al. (2019), Katyal et al. (2019), and Bower et al. (2019). Our model predicts whole mantle melting for $T_{mai} \geq 3000$ K with negligible melt below $T_{mai} \sim 1750$ K. Melting curves differ between experimenters and differ depending on assumed mantle composition (e.g. Andrault et al. 2017, Miyazaki & Korenaga 2019). To relate pressure to depth, we use the pressure-density curve from Dziewonski & Anderson (1981). To get the radius at the magma-atmosphere interface we assume $(R_{pl}/R_\oplus) = (M_{pl}/M_\oplus)^{0.27}$ (Valencia et al. 2006). We integrate downward from the magma-atmosphere interface. Starting at $P_{atm}$ and $T_{mai}$, we follow the adiabat until we reach the solidus. The melt mass is the mass of melt (if any) between the magma-atmosphere interface and the depth of the magma solidus. If this melt mass exceeds 2/3 of planet mass, the melt mass is set equal to 2/3 of planet mass. The effect of the weight of the atmosphere on the location of the solidus is included in our calculations, but turns out to be unimportant.

### *1e. Redistribution of volatiles between atmosphere, magma, and rocks.*

We assume full re-equilibration between magma and atmosphere as the atmosphere is removed (efficient degassing). Degassing is efficient if the cooling magma ocean is fully convective (Ikoma et al. 2018), which can be understood as follows. Whether or not a fully liquid magma ocean undergoing whole-magma-ocean convection will degas as the $H_2$ envelope is removed and the liquid cools depends on (a) the number of times each magma parcel cycles through the upper thermal boundary layer during cooling, and (b) the ratio of the degassed boundary layer thickness to the thermal boundary layer thickness. Suppose the degassed boundary layer thickness is equal to the maximum depth at which bubbles can form (this assumes swift bubble ascent, and that supersaturation is unimportant). For sub-Neptune to super-Earth conversion, this is (up to 10s kbar)/($\rho_{magma}\, g$) = up to 50 km. The thermal boundary layer thickness for a fully liquid magma ocean will be $\ll 50$ km. Each fluid parcel will go through the upper thermal boundary layer at least once during cooling.

We calculate the concentration of dissolved gas in magma using Henry's law. Many gases show linear solubility in magma. Others do not. For example, $H_2O$ dissolves in magma in proportion to the square root of partial pressure at low $pH_2O$, becoming more linear at high $pH_2O$ (e.g. Stolper 1982, Matsui & Abe 1986).

We adopt $s_{H2} = 2 \times 10^{-12}$ Pa$^{-1}$. The basis for this is as follows. Hirschmann et al. (2012) report results from laboratory experiments on basaltic melts. Extrapolating to melts of peridotitic composition, they state that "at 1 GPa in the presence of pure $H_2$, the molecular $H_2$ concentration [in the melt] is 0.19 wt%." Fegley et al. (2020, their Table 5) compile solubilities of various volatiles in molten silicates.

We ignore non-ideal solubility (due to non-ideal fugacity of the gas in the atmosphere), because it is not very important at the relatively low $P_{atm}$ we consider (Kite et al. 2019). We also ignore joint-solubility effects because relevant data are not available. Few measurements have been made of the temperature dependence of volatile solubility in magma at $T > 2000$ K (e.g. Fegley & Schaefer 2014, Guillot & Sator 2011) and we neglect $T$ dependence of volatile solubility.

Solid-melt distribution coefficients, $D$, for water partitioning into nominally anhydrous mantle minerals vary from $10^{-4}$ to $10^{-1}$ (Table S1 in Elkins-Tanton 2008), but for the most common mantle minerals are 0.02 or less. Here, "nominally anhydrous" means minerals that do not have H in their chemical formula. We use $D = 0.02$, which is at the high end of experimentally observed range (corresponding to water portioning into pyroxene). We neglect saturation limits. C distribution coefficients are much lower, from $2 \times 10^{-4}$ to $7 \times 10^{-3}$ according to Elkins-Tanton (2008) and with low (1-5 ppmw) saturation limits.

### *1f. Magma-atmosphere coevolution.*

Key initial conditions are that the quantity of the high molecular weight volatile initially in the crystal phase is zero ($s_{xtl} = 0$), and

$$e_s = C_s/(1 + (\mu_p/\mu_s)(g_{mai}/A_{mai}) m_{magma} s_s) \qquad (S7)$$

which assumes that the atmosphere is dominantly $p$ (i.e., $H_2$) at the zeroth timestep.

We restrict ourselves to $F/F_\oplus < 10^3$. For $F/F_\oplus > 10^3$, silicate vapor-pressure equilibrium atmospheres develop (e.g. Kite et al. 2016).

Our model loops through the following steps: (Step 1) Lose some atmosphere. → (Step 2) Post-escape outgassing. → (Step 3) Crystallize (sequestering some volatiles, but increasing the concentration of volatiles in the magma). → (Step 4) Post-crystallization outgassing → Loop back to Step 1.

<u>Step 1. Lose atmosphere.</u> A down-step in atmospheric pressure is prescribed. No fractionation is permitted between $H_2$ and the high-$\mu$ species during this escape (see section 1g). The $H_2$ and the high-$\mu$ species are lost in proportion to their bulk abundance in the atmosphere.

<u>Step 2. Post-escape outgassing.</u> The magma and atmosphere now re-equilibrate through outgassing of both $H_2$ and $s$ until the partial pressures of $H_2$ and of $s$ are in equilibrium with the concentrations of $H_2$ and $s$ dissolved in the magma. The partial pressure of $H_2$ depends on the mass of $H_2$ in the atmosphere and on the mean molecular weight of the atmosphere, which is affected by the amount of $s$ in the atmosphere. Similarly the partial pressure of $s$ depends on the mass of $s$ in the atmosphere and on the mean molecular weight of the atmosphere, which is affected by the amount of $p$ in the atmosphere. From Eqn. (2) in the main text, we obtain the coupled equations

$$C_s = e_s + e_s((e_s + e_p)/(e_s/\mu_s + e_p/\mu_p)) k_1/\mu_s \tag{S8}$$
$$C_p = e_p + e_p((e_s + e_p)/(e_s/\mu_s + e_p/\mu_p)) k_2/\mu_p \tag{S9}$$

where $C_s$ (kg) is the total on-planet mass of $s$, $C_p$ (kg) is the total on-planet mass of $H_2$ ($p$), $e_s$ is the kg of $s$ in the atmosphere, $e_p$ is the kg of $H_2$ in the atmosphere, the second term on the right-hand-side of (S8) is the mass (kg) of $s$ dissolved in the magma, the second term on the right-hand-side of (S9) is the mass (kg) of $H_2$ dissolved in the magma, $\mu_s$ is the molecular weight of s, $\mu_p$ is the molecular weight of $H_2$, and

$$k_1 = (g_{mai}/A_{mai}) m_{magma} s_s \tag{S10}$$

$$k_2 = (g_{mai}/A_{mai}) m_{magma} s_p \tag{S11}$$

where $g_{mai}$ and $A_{mai}$ are (respectively) the gravity at, and area of, the magma-atmosphere interface for $(R_{mai}/R_\oplus) = (M_{pl}/M_\oplus)^{0.27}$, $m_{magma}$ is calculated as described in section 1d above, $s_s$ is the solubility of $s$ in the magma, and $s_p$ is the solubility of $H_2$ in the magma.

The simultaneous equations (S10)-(S11) have two solutions, and the physical solution is

$$e_p = (C_p k_1 + C_s k_2 + (k_2 (C_p k_1 \mu_s - k_1 (C_p{}^2 k_1{}^2 \mu_p{}^2 + 2 C_p{}^2 k_1 \mu_p \mu_s + C_p{}^2 \mu_s{}^2 + 2 C_p C_s k_1 k_2 \mu_p \mu_s - 2 C_p C_s k_1 \mu_p{}^2 + 4 C_p C_s k_1 \mu_p \mu_s + 4 C_p C_s k_2 \mu_p \mu_s - 2 C_p C_s k_2 \mu_s{}^2 + 2 C_p C_s \mu_p \mu_s + C_s{}^2 k_2{}^2 \mu_s{}^2 + 2 C_s{}^2 k_2 \mu_p \mu_s + C_s{}^2 \mu_p{}^2 )^{0.5} - C_s k_1 \mu_p + 2 C_s k_2 \mu_s + C_p k_1{}^2 \mu_p + C_s k_1 k_2 \mu_s ))/(2 (k_1 \mu_p - k_2 \mu_s + k_1{}^2 \mu_p - k_1 k_2 \mu_s )) + (k_1 k_2 (C_p k_1 \mu_s - k_1 (C_p{}^2 k_1{}^2 \mu_p{}^2 + 2 C_p{}^2 k_1 \mu_p \mu_s + C_p{}^2 \mu_s{}^2 + 2 C_p C_s k_1 k_2 \mu_p \mu_s - 2 C_p C_s k_1 \mu_p{}^2 + 4 C_p C_s k_1 \mu_p \mu_s + 4 C_p C_s k_2 \mu_p \mu_s - 2 C_p C_s k_2 \mu_s{}^2 + 2 C_p C_s \mu_p \mu_s + C_s{}^2 k_2{}^2 \mu_s{}^2 + 2 C_s{}^2 k_2 \mu_p \mu_s + C_s{}^2 \mu_p{}^2 )^{0.5} - C_s k_1 \mu_p + 2 C_s k_2 \mu_s + C_p k_1{}^2 \mu_p + C_s k_1 k_2 \mu_s ))/(2 (k_1 \mu_p - k_2 \mu_s + k_1{}^2 \mu_p - k_1 k_2 \mu_s )))/(k_1 + k_1 k_2) \tag{S12}$$

$$e_s = -(C_p k_1 \mu_s - k_1 (C_p{}^2 k_1{}^2 \mu_p{}^2 + 2 C_p{}^2 k_1 \mu_p \mu_s + C_p{}^2 \mu_s{}^2 + 2 C_p C_s k_1 k_2 \mu_p \mu_p - 2 C_p C_s k_1 \mu_p{}^2 + 4 C_p C_s k_1 \mu_p \mu_s + 4 C_p C_s k_2 \mu_p \mu_s - 2 C_p C_s k_2 \mu_s{}^2 + 2 C_p C_s \mu_p \mu_s + C_s{}^2 k_2{}^2 \mu_s{}^2 + 2 C_s{}^2 k_2 \mu_p \mu_s + C_s{}^2 \mu_p{}^2 )^{0.5} - C_s k_1 \mu_p + 2 C_s k_2 \mu_s + C_p k_1{}^2 \mu_p + C_s k_1 k_2 \mu_s )/(2 (k_1 \mu_p - k_2 \mu_s + k_1{}^2 \mu_p - k_1 k_2 \mu_s )) \tag{S13}$$

We recalculate the other variables by inserting the new $e_s$ and $e_p$ into Eqns. (S8) and (S9).

The coupling of Eqn. (S8) and Eqn. (S9) via $\mu_{avg}$ has two important effects that disfavor secondary atmospheres on exoplanets. During the loss of the primary atmosphere, the $\mu_{avg}$ effect increases the amount of the high-$\mu$ constituent in the atmosphere by a factor of 15 (= $\mu_s/\mu_p$), increasing the amount of the high-$\mu$ constituent that is lost to space. Moreover, as $\mu_{avg}$ rises from 2 to 30, because the partial pressure of the high-$\mu$ constituent scales with the number ratio of $s$ in the atmosphere, $f_s = e_s / (e_p + e_s)(\mu_{avg}/\mu_p)$, the number of moles of $s$ in the atmosphere for a given saturation vapor pressure of $s$ decreases by a factor of 15 ( = $\mu_s/\mu_p$). This negative feedback reduces the number of worlds that smoothly transition from a primary to a secondary atmosphere.

<u>Step 3. Crystallize.</u> The magma mass is forced to a lower value, consistent with the lower $T_{mai}$ associated with the atmospheric pressure that was reduced in Step 1. The freshly-produced crystals gain an $s$ concentration (in mass fraction) of $D_s \times X_s$, where $X_s$ is the mass fraction of the volatile in the magma. The crystals are not remixed (by assumption) on the timescale of magma-ocean crystallization. $s$ within crystals is relatively safe from release into the unsafe surface environment during sub-Neptune-to-super-Earth conversion, because solid-state creep speeds within the solid mantle are $10^{-9} - 10^{-8}$ m s$^{-1}$, versus up to $10^1$ m s$^{-1}$ in the magma. As a result, the concentration of the crystals is independent of the previous partitioning of the volatiles into the solid rock. Partitioning of $H_2$ into the crystals is not considered.

<u>Step 4. Post-crystallization outgassing.</u> Because $D_s < 1$, the magma ocean is enriched in $s$ after the crystallization step. Therefore we re-equilibrate the magma and the atmosphere taking account of the now-higher concentration of $s$ ($X_s$) in the magma.

The processes involved in Steps 1-4 all occur on timescales short compared to the magma ocean lifetime.

## *1g. Calculating evolution over time.*

To map the rocky planet evolution sequence onto time we combine the environmental drivers from (1a), the loss rates from (1b), and the small planet evolution sequence from (1f). $T_{RCB}$ is fixed in the crystallization calculation, but $T_{eq}$ evolves with star age. We deal with this by setting $T_{RCB}$ throughout the calculation equal to $T_{eq}$ at a fixed age of 1 Gyr. Thus, planets orbiting cool stars in our model have shorter-lived magma oceans than in reality.

Diffusive separation of $H_2$ from high-$\mu$ species is important at lower XUV flux (e.g. Hu et al. 2015, Wordsworth et al. 2018, Saito & Kuramoto 2018, Odert et al. 2018, Malsky & Rogers 2020). However, diffusive separation is washed out at the high XUV fluxes required if photoevaporation is to cause sub-Neptune-to-super-Earth conversion (Tian 2015, Catling & Kasting 2017). To demonstrate this, we first define the (atomic) H flux, $F_1$, on the assumption that the overwhelming majority of the upper atmosphere is comprised of H:

$$F_1 = \varepsilon\, F_{XUV}\, (R + z) / (4\, G\, M_{pl}\, m_1) \tag{S14}$$

in units of molecules s$^{-1}$ m$^{-2}$. Here, $\varepsilon = 0.15$ is the efficiency of conversion of XUV energy into atmospheric escape, $R$ is solid-planet radius, $z$ is atmosphere thickness (which we set to $0.4\,R$ to correspond to a small sub-Neptune), $G$ is the gravitational constant, $M_{pl}$ is planet mass, and $m_1$ is the mass (kg) of the hydrogen atom.

From this we obtain the crossover mass $m_c$ (Tian 2015)

$$m_c = m_1 + k\,T\,F_1 / (b\,g\,X_1) \tag{S15}$$

where $T$ is the temperature of the heterosphere, and $b = 4.8 \times 10^{19}\,T^{0.75}$ m$^{-1}$ s$^{-1}$ is the bindary diffusion coefficient, corresponding to neutral H and neutral O, used by Tian (2015). Including the effect of ionization of H would decrease the effective value of $b$ (Hu et al. 2015) and so strengthen our conclusion that diffusive separation is not important during sub-Neptune to super-Earth conversion for *Kepler*'s super-Earths. We set $T = 10^4$ K, the same value used by Hu et al. (2015) and Malsky & Rogers (2020).

The flux, $F_2$, of the high-$\mu$ species is zero when $m_c < m_2$. When $m_c > m_2$ the flux $F_2$ is given by

$$F_2 = F_1\,(X_2\,(m_c - m_2)) / (X_1\,(m_c - m_1)) \tag{S16}$$

Substituting in values for 6 $M_\oplus$ and atomic mass 15 Da we find $F_{XUV} = 0.7$ W/m$^2$ at crossover, 1.3 W/m$^2$ for 50% reduction in escaping flux due to weight differences, and 6.1 W/m$^2$ for escaping flux of high-molecular weight species at 90% of the flux if there was no fractionation at all. All three thresholds are shown as dashed vertical lines on Fig. S5 and Fig. S6. In the context of sub-Neptune to super-Earth conversion, most of the atmosphere mass is removed at higher $F_{XUV}$. Most exoplanets either stay as sub-Neptunes or lose their atmosphere too quickly for the XUV flux to drop to levels that permit >10% differences in loss rate due to fractionation. This can be understood as follows. If the upper atmosphere absorbs 1 W/m$^2$ of light from the star, the upper limit on loss rate for a 6 M$_\oplus$ world is ~10 bars/Myr, falling to 1.5 bars/Myr at $\varepsilon = 0.15$. Since sub-Neptune-to-super-Earth conversion takes ~300 Myr, and ~5 × 10$^4$ bars of H$_2$ need to be removed (including H$_2$ that is initially dissolved in the magma), $F_{XUV}$ must be ~100 W/m$^2$. Using the saturation threshold of Selsis et al. (2007), ($F_{XUV}/F = 10^{-3.2}$) this corresponds to 160 kW/m$^2$, $F/F_\oplus = 120$, for distance from the star = 0.1 AU for a Sun-like star. This is the typical distance of a Kepler super-Earth from its host star (the abundance of 1.6 $R_\oplus \sim$ 6 $M_\oplus$ super-Earths drops off rapidly at larger separations (Fulton et al. 2017, Fulton & Petigura 2018). At such high XUV fluxes, fractionation causes negligible differences in the loss rate relative to the no-fractionation case. Therefore, our approximation of no fractionation during sub-Neptune to super-Earth conversion is valid.

However, fractionation by diffusion can be important for the composition of late-stage volcanically outgassed atmospheres. For example, the volcanic H$_2$ outgassing flux is calculated to be less than the diffusively-limited H$_2$ escape rate unless H$_2$ stays at a very low level (e.g. Catling et al. 2001, Batalha et al. 2016, Zahnle et al. 2019). We expect that this effect would maintain volcanically outgassed atmospheres at high $\mu_{avg}$, even if outgassing of H$_2$ by serpentinization (Sleep et al. 2004) and/or volcanic outgassing is large.

Fractionation by diffusion is also much more important for Habitable Zone super-Earth-sized planets. For $F_{bol}$ = 1361 W/m$^2$ (the Solar flux at 1 AU today), even at saturation $F_{XUV}$ is 0.9 W/m$^2$, which is only marginally capable of lifting atomic mass = 15 Da species out of the atmosphere. This favors the occurrence of secondary atmospheres on Habitable Zone exoplanets.

Finally, exposure to moderate $F_{XUV}$ for Gyr may convert a $H_2$-dominated sub-Neptune atmosphere into a He-enriched sub-Neptune, provided the planet does not start with too much $H_2$ (Hu et al. 2015, Malsky & Rogers 2020).

### *1h. Volcanic outgassing.*

No exoplanets have yet been confirmed to be volcanically active. We use a simple model of the geodynamics and rate of volcanism on super-Earth-sized planets (Kite et al. 2009). The model includes parameterized mantle convection, thermal evolution, and volcanism for both plate tectonics and stagnant lid modes. The model is tuned to reproduce Earth's present-day rate of volcanism in plate tectonics mode. Kite et al. (2009) consider three different melting models. In the present study, we use the results from the widely-used melting model of Katz et al. (2003). Many more sophisticated models of rocky exoplanet thermal evolution, geodynamics, and volcanic activity exist (e.g. Dorn et al. 2018, and references therein).

The procedure for estimating the rate of volcanism starts from the results of Kite et al. (2009) (Fig. S3). This is an Earth-tuned parameterized mantle convection and rate of volcanism model that predicts the rate of volcanism versus time for either stagnant lid mode and plate tectonics mode. The rate of $CO_2$ release at Earth's mid-ocean ridges (12±2 bars/Gyr; Tucker et al. 2018) is multiplied by the computed rate of volcanism, in units of Earth's present-day rate, to obtain the predicted rate of $CO_2$ release on the rocky exoplanet. This is adjusted downward in proportion to any early loss of high-molecular-weight volatiles to space. For example, if a model exoplanet has lost 70% of its high-molecular-weight volatiles to space, then the rate of volcanic outgassing is reduced to 30% of the volumetric volcanic flux predicted from the Earth-tuned model of Super-Earth volcanism.

Fig. 6 compares the broad zone of worlds with atmosphere for planets that form without thick $H_2$ atmospheres to the much more restricted zone of volcanically revived atmospheres for worlds that form with thick $H_2$ atmospheres. This calculation assumes that half of worlds form with $s_s$ = $10^{-9}$ Pa$^{-1}$ (for which ~90% of high-$\mu$ volatiles are lost to space during the loss of the $H_2$-dominated atmosphere), and half form with $s_s$ = $10^{-11}$ Pa$^{-1}$. Atmospheric ingassing is neglected because worlds well inside the habitable zone are too hot for weathering reactions to sequester atmophiles in the crustal rocks. Volatile release by volcanic outgassing is neglected, by construction, on worlds that start with all their volatiles in the atmosphere. Forming with a secondary atmosphere is advantageous for subsequently exhibiting a secondary atmosphere, relative to the total-loss-and-subsequent-revival process. The stagnant-lid and plate-tectonics predictions diverge greatly after ~4 Gyr, when volcanism shuts down on the stagnant-lid model planet (Fig. S8). The plate-tectonics and stagnant lid predictions otherwise appear similar in this Earth-scaled model (Kite et al. 2009). However, this overall similarity of plate and stagnant-lid predictions is specific to the Earth-tuned model of Kite et al. (2009): at least one other model predicts much stronger suppression of volcanism on stagnant-lid super-Earth-sized rocky planets (e.g. Dorn et al. 2018). Tidal locking, by itself, has little effect on volcanism (e.g. van Summeren et al. 2011). So long as we do not mechanistically understand volcanism versus time for Solar System worlds (Byrne et al. 2019), why Earth's mantle took so long to mix, nor why Earth's mantle is not yet completely outgassed, we think that it is appropriate to use basic models to predict rate of volcanism versus time for rocky exoplanets. A robust constraint (for non-tidally-heated planets) is decline of mantle temperature by 50-200K per Gyr, with a concomitant decline in the potential for melting (Stevenson 2003).

We assume the rate of degassing is proportional to the rate of magmatism. This is a simplification. Even low-degree partial melts can effectively extract almost all the volatiles from the full mass of silicate mantle that sweeps through the partial melt zone. This is because volatiles partition strongly into the melt during partial melting (i.e., $D_i \ll 1$). Once the partial melt fraction is high enough (~1%) for the melt to form interconnected channels, the volatile bearing melt rises from the melt production zone geologically instantaneously (with no loss of volatiles) via filter-pressing, melt channel formation, and diking. Whether the melt gets close enough to the surface to form bubbles and/or erupt explosively will depend on the stress state of the lithosphere (e.g. Solomon 1978). Explosive volcanic eruptions are aided by bubble formation, which is easier when the volatile content of the magma is high. If melt crystallizes at depth (intrusion) then volatiles will go into hydrated minerals or into the glass phase, where they might still be released to the atmosphere over geologic time.

Extrapolation of simple models of Earth's thermal evolution and outgassing rate back into Earth's past leads to the expectation that Earth's mantle was very rapidly stirred and quickly outgassed very early in Earth history – in contradiction to isotopic data that show sluggish stirring. Similarly, our simple Earth-tuned model of thermal evolution and outgassing rate indicates that 4-6× more volatiles outgas over the lifetime of the planet than are present at the beginning – a mass balance violation. Because an excess of volcanism over XUV-driven atmospheric loss leads (in our model) to build-up of a detectable atmosphere in <1 Gyr, the rate of volcanism is more important than the cumulative amount of volcanism in setting atmosphere presence/absence. Therefore we do not adjust our volcanic fluxes downward to take account of this effect, reasoning that real planets outgas more slowly than simple models predict. Alternatively, our results could be interpreted as saying that super-Earths erupt away their solid-mantle volatiles during the first Gyr of the star's life, when XUV fluxes are still high. This would lead to an even more unfavorable conclusion for the likelihood of volcanically outgassed atmospheres at distances much closer to the star than the habitable zone than is presented in Fig. 6.

To calculate the atmosphere presence/absence lines shown in Fig. 6 and Fig. S16, we varied following parameters. (1) XUV flux: multiplied by a factor drawn randomly from a lognormal distribution centered on 1 and with a standard deviation of 0.4 dex (Lyon et al. 2019). (2) Earth $CO_2$ outgassing rate: a quantity drawn randomly from a normal distribution centered on 0 and with a standard deviation of 2 bars/Gyr is added to the nominal value of 12 bars/Gyr. (3) Solubility of the high-molecular-weight species in the magma: alternated between values of $10^{-11}$ Pa$^{-1}$ and $10^{-9}$ Pa$^{-1}$. We used 1 bar atmospheric pressure as the threshold for atmosphere presence/absence.

The model results are more sensitive to changes in the XUV flux than they are to changes in the rate of volcanic outgassing. This is due to the exponential cutoff in the XUV-flux-driven loss rate of $CO_2$ (Tian 2009).

## 1i. Supply and observability.

The inventory of $CO_2$ even on today's Earth is uncertain. For example, the review of Lee et al. (2019) reports a range of estimates for the C stored in the non-sedimentary rocks of Earth's continental crust, from $4.2 \times 10^7$ Gton C to $2.6 \times 10^8$ Gton C. This increases the likelihood that some hot rocky exoplanets have C inventories large enough to retain an observable atmosphere.

No exoplanets have yet been confirmed to contain >1 wt% $H_2O$. Modeling papers tracking the formation, migration, and (for habitable-zone worlds) climate evolution of worlds with >1 wt% $H_2O$ include Kite & Ford (2018), Bitsch et al. (2019), and references therein.

Another way to increase volatile mass is to ballast the hydrogen with oxygen obtained from the reaction $FeO_{(magma)} + H_{2(g)} \rightarrow Fe + H_2O_{(g)}$; i.e., endogenic water which does not require planet migration (Kite et al. 2020).

**SI References.**

Wordsworth, R.D., et al. 2018, Redox evolution via gravitational differentiation on low-mass planets: Implications for abiotic oxygen, water loss, and habitability, Astromical Journal, 155:195.

Zahnle, K., & D. Catling 2017, The cosmic shoreline: The evidence that escape determines which planets have atmospheres, and what this may mean for Proxima Centauri b, Astrophysical Journal, 843:122

Zahnle, K., et al. 2019, Strange messenger: A new history of hydrogen on Earth, as told by Xenon, Geochim. Cosmichim. Acta 244, 56-85.38

**SI Figures.**

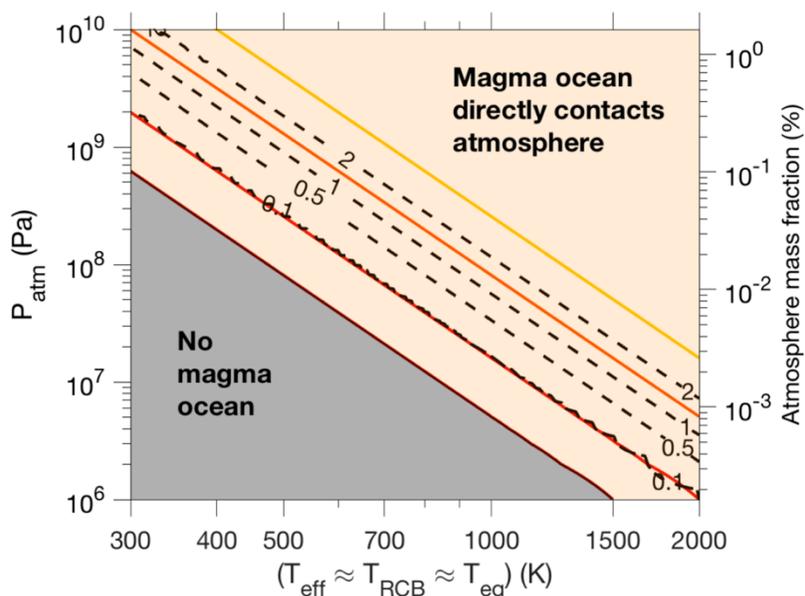

**Fig. S1.** How magma ocean mass increases with atmospheric thickness. Output from a toy model of sub-Neptune thermal structure (Kite et al. 2020). Dashed lines correspond to magma ocean mass, labeled in Earth-masses of magma, for a volatile-free planet mass of 6 $M_\oplus$. Colored lines correspond to temperatures at the magma-atmosphere interface of 1500 K (maroon), 2000 K (red), 3000 K (orange), and 4000 K (yellow). Magma ocean masses in excess of 2 Earth masses are not plotted because this corresponds to a magma pressure range that is not well explored by experiments.

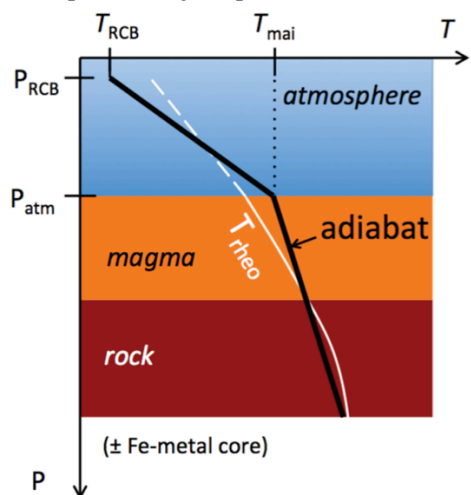

**Fig. S2.** Planet thermal structure (reproduced from Kite et al., 2020). Temperature versus pressure plot, showing adiabats within the atmosphere and the magma (black line). $T_{rheo}$ (white line) corresponds to the temperature of the rheological transition (~40% melt fraction) for rock (dashed white line at levels where no rock is present). *RCB* = radiative-convective boundary. $T_{mai}$ = temperature at the magma-atmosphere interface. For real planets the equilibrium temperature ($T_{eq}$), the effective temperature ($T_{eff}$), and the temperature at the RCB ($T_{RCB}$), are related by $T_{eq} < T_{eff} \lesssim T_{RCB}$. In this paper we make the approximation that these three temperatures are equal (SI Appendix, section 1c).

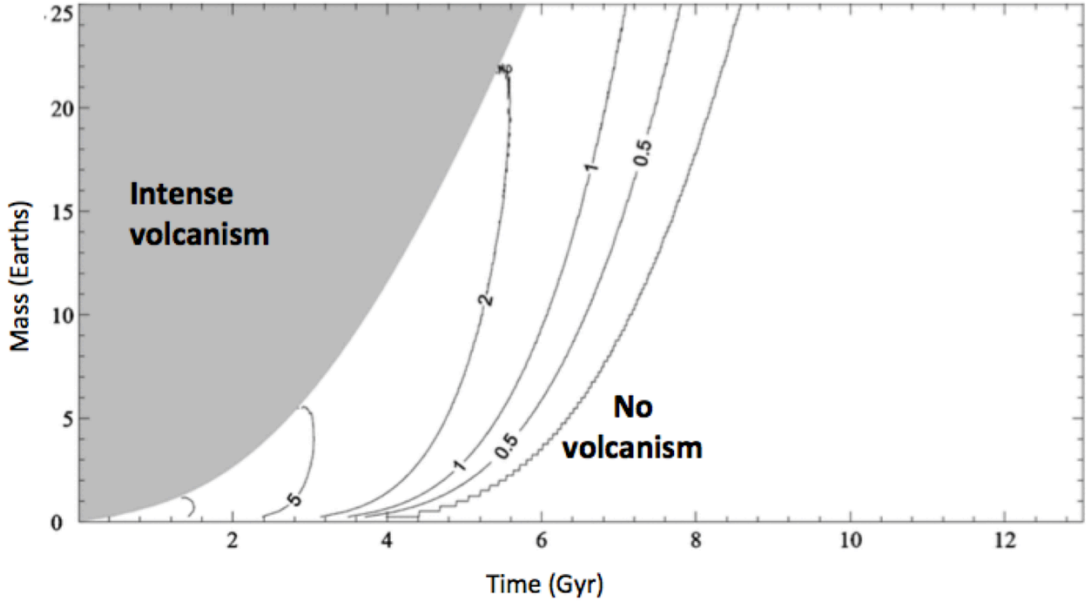

**Fig. S3.** (Reproduced from Kite et al. 2009, ref. 34 in the main text, by permission of the AAS). Rate of volcanism per unit mass on massive Earth-like planets undergoing stagnant lid convection, normalized to calculated rate on a plate-tectonic Earth, for Katz et al. (2003) melting model. Dark gray shaded regions correspond to mantle temperatures associated with very intense volcanism, too high for a reliable crustal thickness calculation. Contours are at 0, 0.5, 1, 2, 5, and 10 times Earth's present-day rate of volcanism ($\sim 4 \times 10^{-19}$ s$^{-1}$, equivalent to 24 km$^3$ yr$^{-1}$, in plate-tectonics mode). For example, a 6 $M_\oplus$ planet on the "2" contour erupts $6 \times 2 \times 24 \approx 300$ km$^3$ yr$^{-1}$.

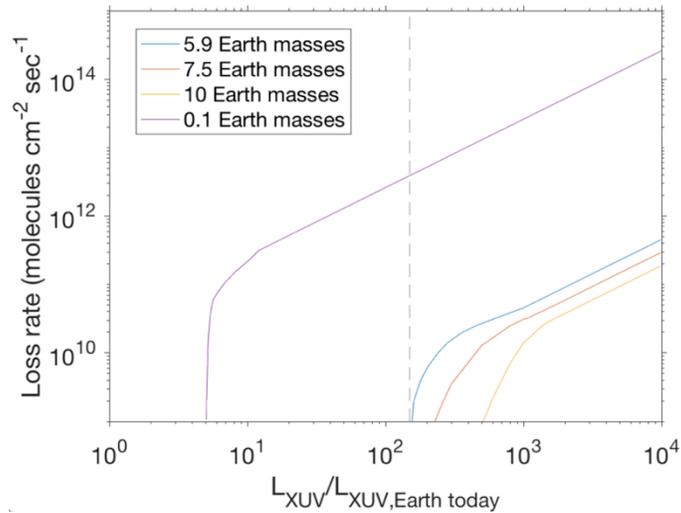

**Fig. S4.** Atmospheric loss rate for a pure $CO_2$ atmosphere, combining results for super-Earths (Tian 2009) and for Mars (Tian et al. 2009, scaled to 1 AU). The vertical dashed line corresponds to 150× Earth's present-day XUV flux.

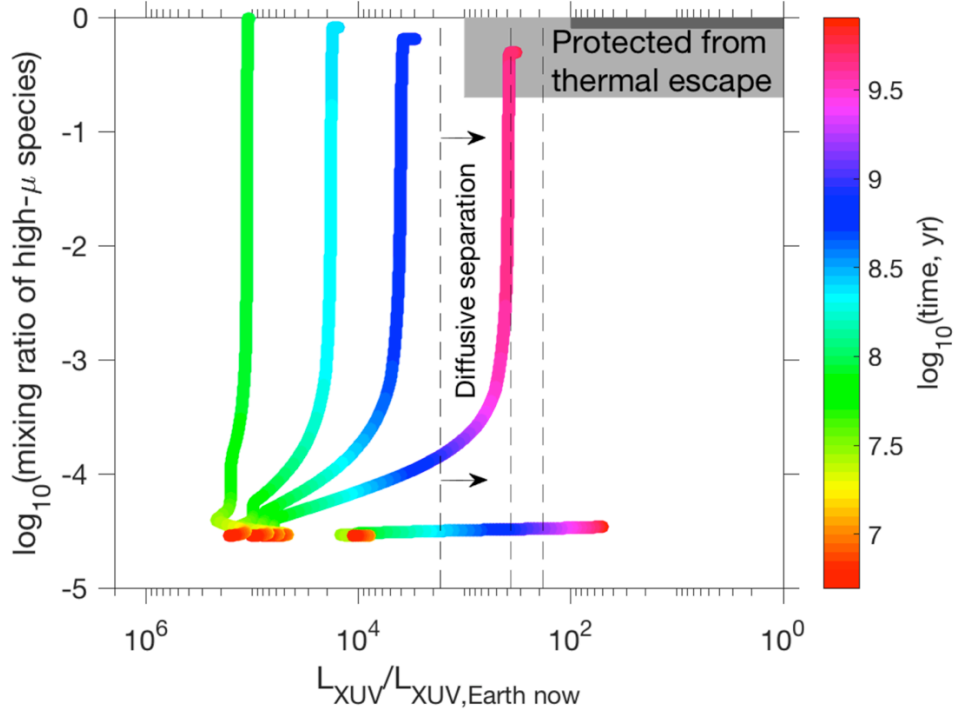

**Fig. S5.** The same evolutionary tracks as in Fig. 5 (main text), shown in $F_{XUV} - \mu_{avg}$ coordinates. $F/F_\oplus$ increases from rightmost track ($F/F_\oplus = 49$, $T_{eq} = 735$ K) to leftmost track ($F/F_\oplus = 720$, $T_{eq} = 1440$ K). The dashed lines highlight (from right to left) the threshold of $F_{XUV}$ below which all of the high-molecular-mass species is retained by the planet; the $F_{XUV}$ corresponding to a 50% reduction in the no-fractionation loss rate of the high-molecular-mass species; and the $F_{XUV}$ corresponding to escape of the high-molecular-mass species at 90% of the rate at which no loss would occur (SI Appendix, section 1g). Right of the rightmost dashed line, fractionation protects the constituents of the secondary atmosphere, and left of the leftmost dashed line, fractionation is much less important. Calculations are done assuming an atomic wind of H entraining a 1% mixing ratio of atoms of mass 15 Da.

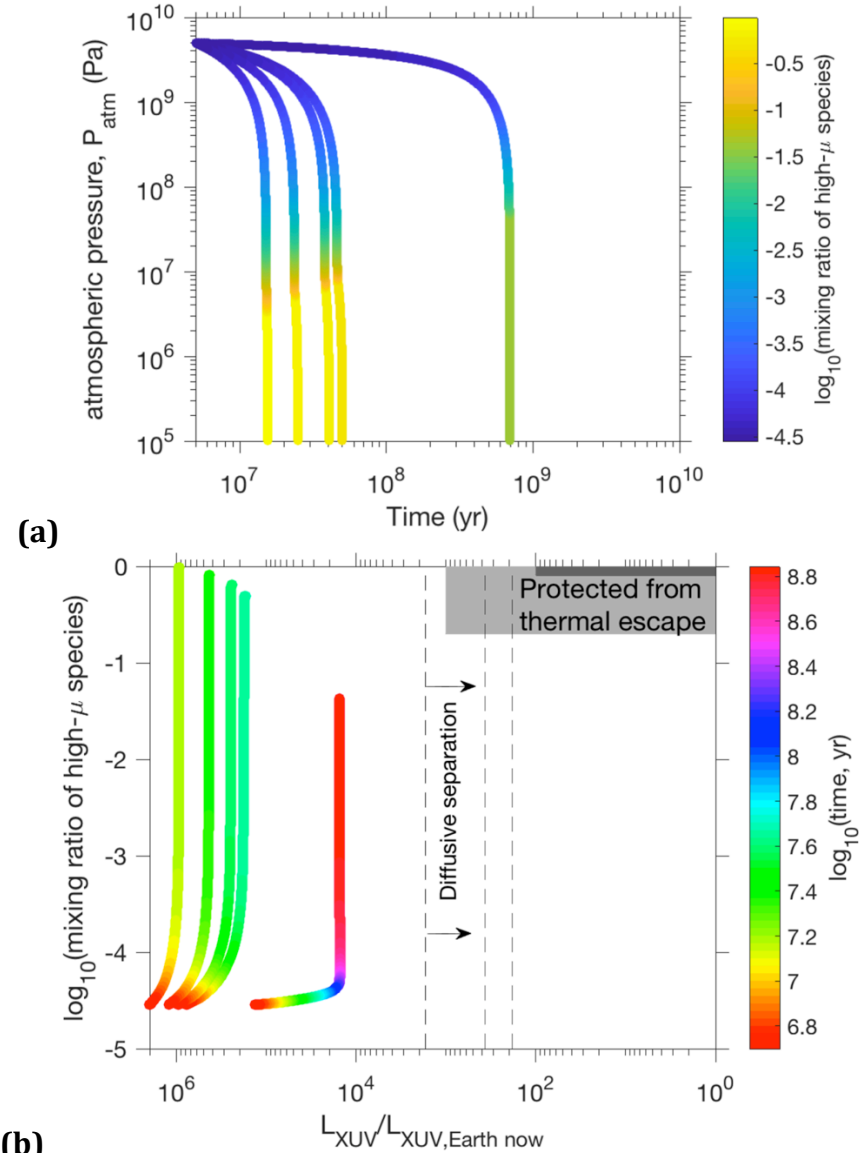

**Fig. S6. (a)** As Fig. 5a (main text), but for a planet orbiting a star of 0.3 Solar masses. Time-dependent results. Atmospheric pressure vs. time for planets for (from right to left) $F/F_\oplus$ = {49, 283, 347, 422, 720}, corresponding to planet equilibrium temperature ($T_{eq}$) = {735, 1140, 1200, 1275, 1440} K. **(b)** As Fig. S5, but for a planet orbiting a star of 0.3 Solar masses. The same evolutionary tracks as in the top panel, shown in $F_{XUV}$– $\mu_{avg}$ coordinates. $F/F_\oplus$ increases from rightmost track ($F/F_\oplus$ = 49, $T_{eq}$ = 735 K) to leftmost track ($F/F_\oplus$ = 720, $T_{eq}$ = 1440 K). The dashed lines highlight (from right to left) the threshold of $F_{XUV}$ below which all of the high-molecular-mass species is retained by the planet; the $F_{XUV}$ corresponding to a 50% reduction in the no-fractionation loss rate of the high-molecular-mass species; and the $F_{XUV}$ corresponding to escape of the high-molecular-mass species at 90% of the rate at which no loss would occur (SI Appendix, section 1g). Right of the rightmost dashed line, fractionation protects the constituents of the secondary atmosphere, and left of the leftmost dashed line, fractionation is much less important. Fractionation calculations are done assuming an atomic wind of H entraining a 1% mixing ratio of atoms of mass 15 Da.

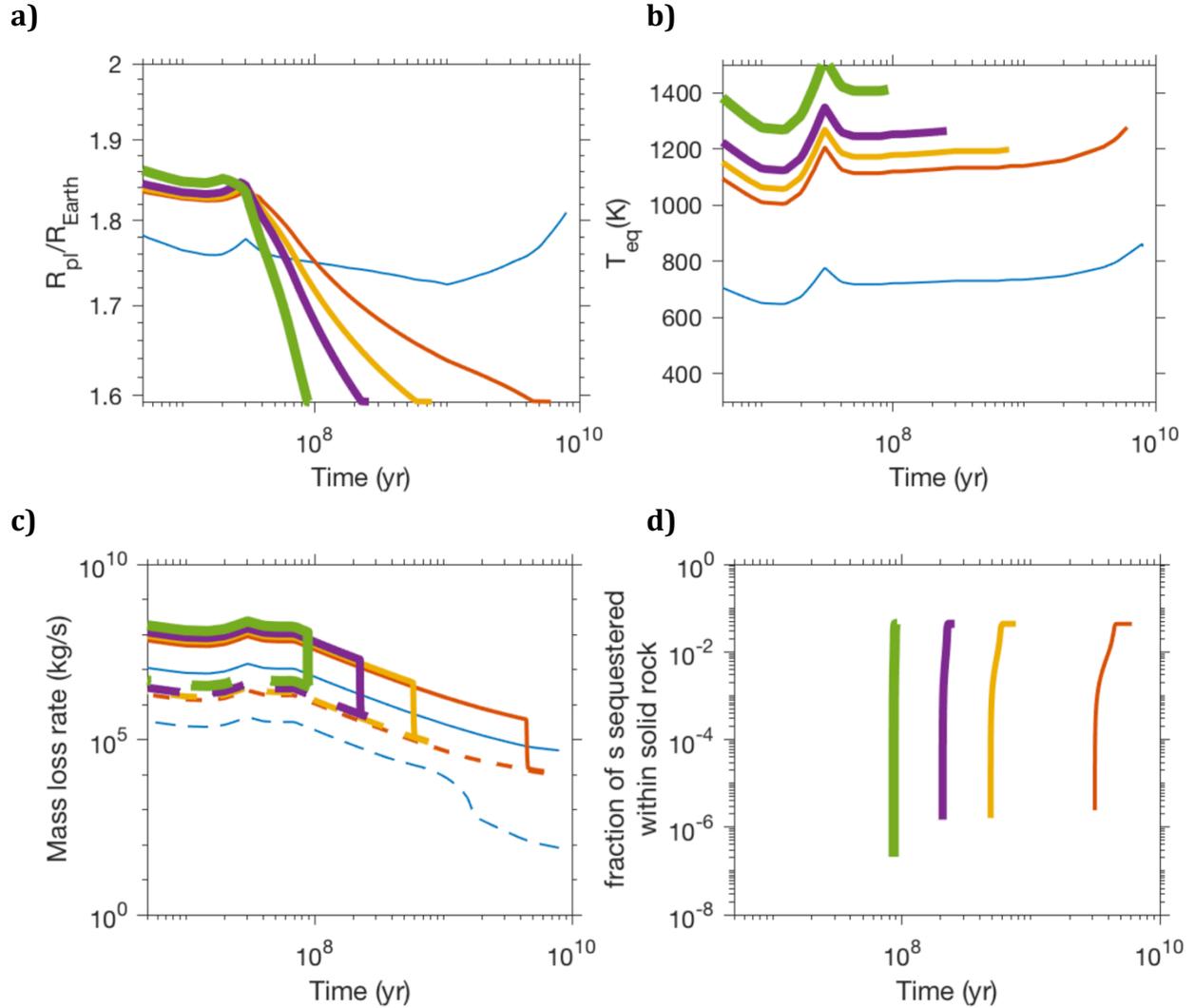

**Fig. S7.** Details of planet evolution for the tracks shown in Fig. 5a (main text). Solar-mass star, 6 $M_\oplus$. Line thickness corresponds to insolation, with the thickest lines corresponding to the greatest insolation. Results are shown for $F/F_\oplus$ = 49 (blue), $F/F_\oplus$ = 283 (red), $F/F_\oplus$ = 347 (yellow), $F/F_\oplus$ = 422 (purple), and $F/F_\oplus$ = 720 (green), corresponding to $T_{eq}$ = {735, 1140, 1200, 1275, 1440} K, respectively. **(a)** Planet radius versus time. **(b)** Planet equilibrium temperature versus time. **(c)** Planet mass loss rate versus time (solid lines), and the mass loss rate that the planet would have if the atmosphere was composed entirely of $CO_2$ based on the calculations of Tian (2009) (dashed lines). The planets corresponding to the red, yellow, purple, and green lines evolve to a high-$\mu$ atmosphere, but only in the red case is this atmosphere long-lived. **(d)** The fraction of the high-$\mu$ species ($s$) that is shielded within solid rock. This is zero for the world that remains as a sub-Neptune (because no solid rock forms), and very similar for the four worlds that transition to super-Earths (full magma crystallization).

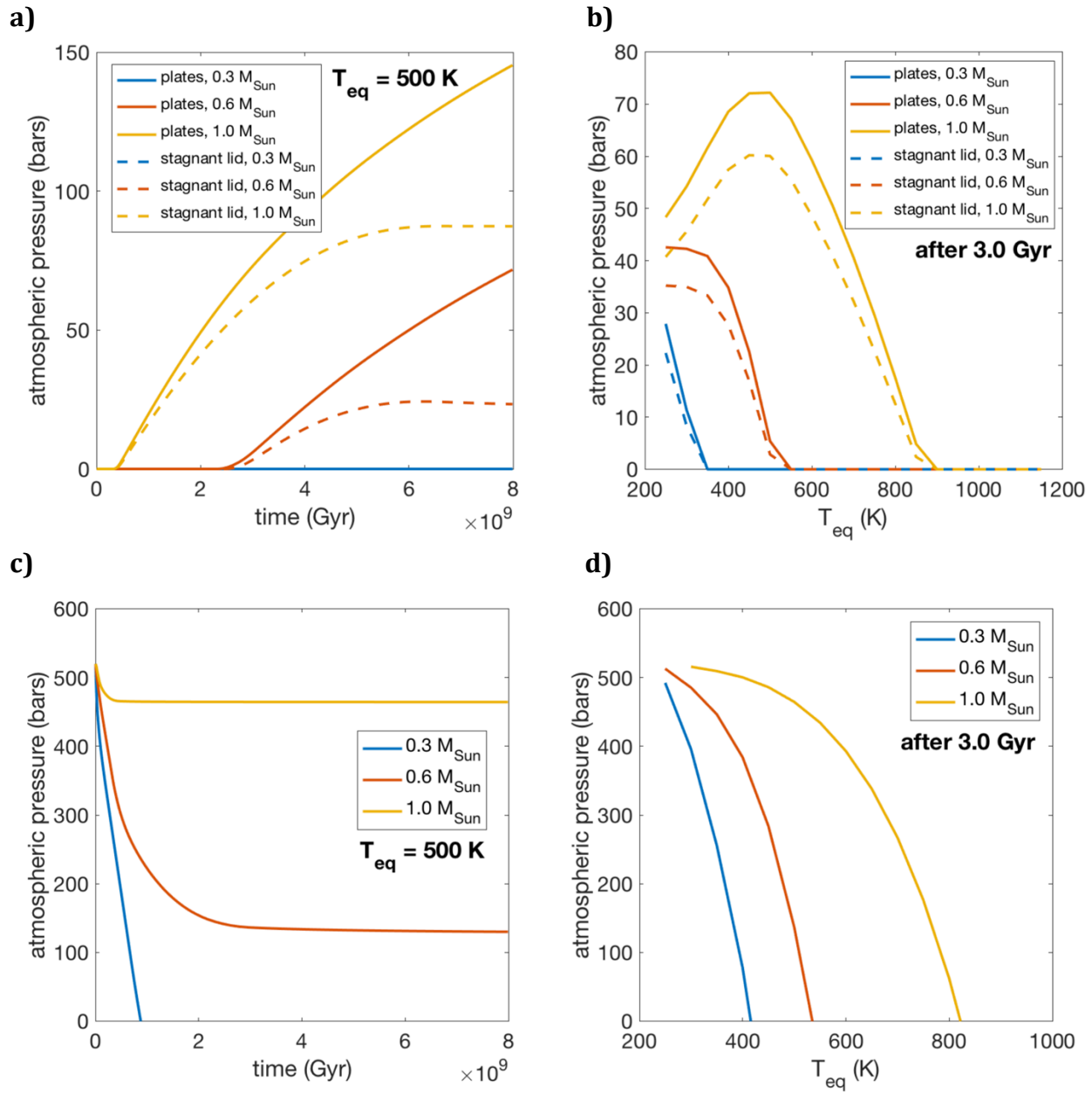

**Fig. S8.** Atmospheric thickness versus time for the Selsis et al. (2007) XUV model. **(a-b)** volcanic revival, $s_s = 10^{-9}$ Pa$^{-1}$, Kite et al. (2009) volcanism model. Results are shown for both plate tectonics (plates), and for stagnant lid mode. There is a minor trend in plate-tectonics mode for planets around Solar-mass stars at $T_{eq} < 500$ K toward decreasing atmospheric thickness with decreasing $T_{eq}$. This is due to the smaller initial volume of the ($s$-protecting) magma ocean at low $T_{eq}$. **(c-d)** Start-with-all-volatiles-in-the-atmosphere, residual atmosphere case. In this model, the planet forms with no H$_2$ (intrinsically rocky) and with 500 bars of $s$, all in the atmosphere.

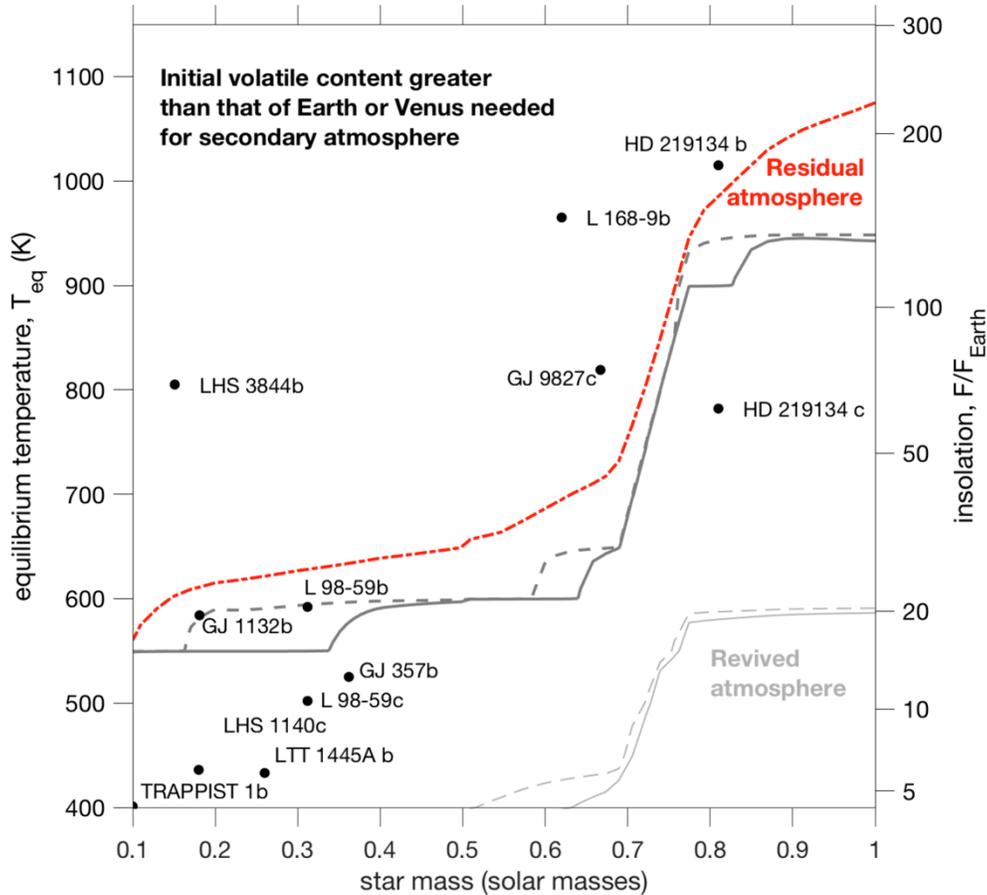

**Fig. S9.** Deterministic calculation for atmosphere presence/absence after 3.0 Gyr. Deterministic calculation with a fixed XUV flux for each star mass, combining the fits of Jackson et al. (2012) and Guinan et al. (2016). The red dash-dot line corresponds to the line of vanished atmospheres for planets that have all volatiles in the atmosphere initially. Worlds at hotter $T_{eq}$ than this line can retain an atmosphere if there is no $H_2$ initially and all high-molecular-weight volatiles are in the atmosphere initially. This line moves away from the star over time. The thick gray solid (plate tectonics) and dashed (stagnant lid) lines correspond to the line of atmospheric revival by volcanic outgassing for planets that lose all atmosphere during transition from a sub-Neptune to a super-Earth. These lines of atmospheric revival sweep towards the star over time because the rate of volcanic degassing falls off more slowly with time than does the star's XUV flux. This chart assumes initial volatile supply is independent of $F$ and similar to Earth and Venus. Increasing volatile supply will move thresholds to higher $F$. The thin gray lines show results for a reduction in the solubility of the high-molecular-weight volatile in the magma from $10^{-9}$ Pa$^{-1}$ to $10^{-11}$ Pa$^{-1}$. Selected exoplanets overplotted. Note that because most of the planets that are shown are too small to have Earth-like composition and 6 $M_\oplus$, this plot is optimistic for atmospheric survival (higher planet mass favors atmosphere retention). For any individual planet, star-specific XUV-flux estimates, star age, and the planet's mass, should be combined to make a more accurate estimate than is possible using this overview diagram.

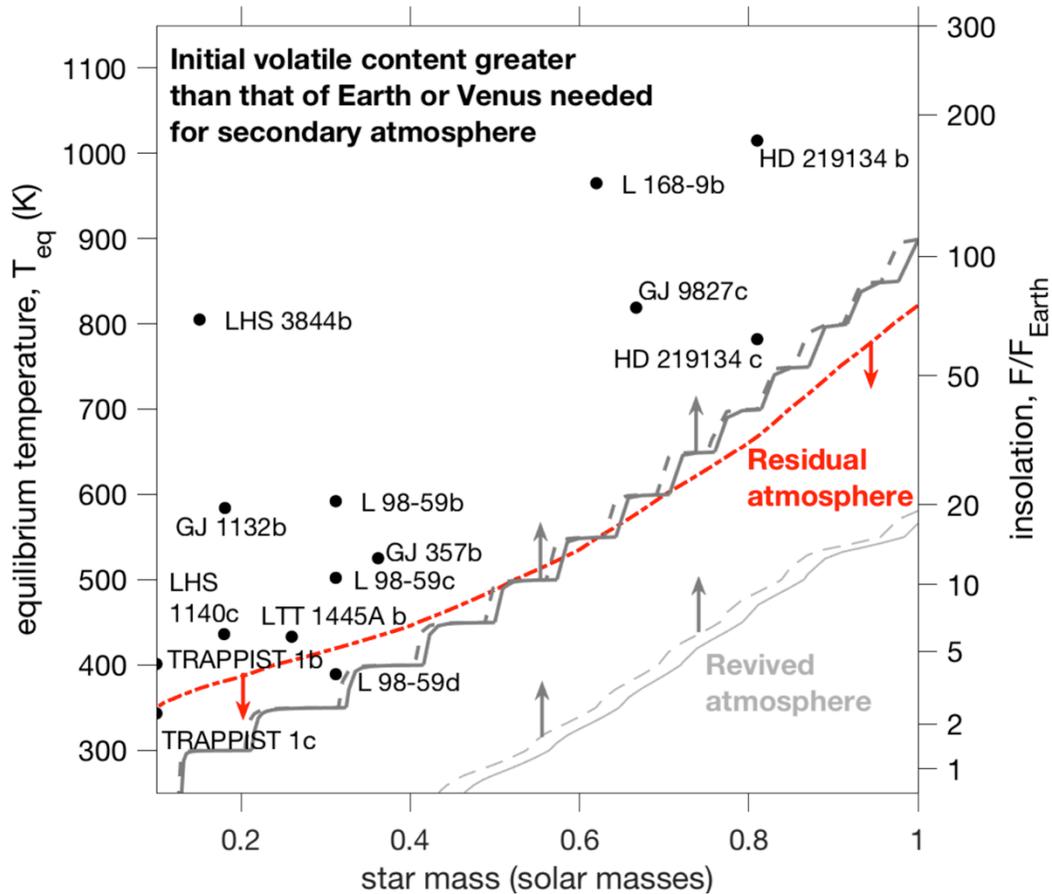

**Fig. S10.** Deterministic calculation for atmosphere presence/absence after 3.0 Gyr. Deterministic calculation with a fixed XUV flux for each star mass, following Selsis et al. (2007). The red dash-dot line corresponds to the line of vanished atmospheres for planets that have all volatiles in the atmosphere initially. Worlds at hotter $T_{eq}$ than this line can retain an atmosphere if there is no $H_2$ initially and all high-molecular-weight volatiles are in the atmosphere initially. This line moves away from the star over time (red arrows). The thick gray solid (plate tectonics) and dashed (stagnant lid) lines correspond to the line of atmospheric revival by volcanic outgassing for planets that lose all atmosphere during transition from a sub-Neptune to a super-Earth. These lines of atmospheric revival sweep towards the star over time (gray arrows) because the rate of volcanic degassing falls off more slowly with time than does the star's XUV flux. This chart assumes initial volatile supply is independent of $F$ and similar to Earth and Venus. Increasing volatile supply will move thresholds to higher $F$. The thin gray lines show results for a reduction in the solubility of the high-molecular-weight volatile in the magma from $10^{-9}$ Pa$^{-1}$ to $10^{-11}$ Pa$^{-1}$. Selected exoplanets overplotted. Note that because most of the planets that are shown are too small to have Earth-like composition and $6\,M_\oplus$, this plot is optimistic for atmospheric survival (higher planet mass favors atmosphere retention). For any individual planet, star-specific XUV-flux estimates, star age, and the planet's mass, should be combined to make a more accurate estimate than is possible using this overview diagram.

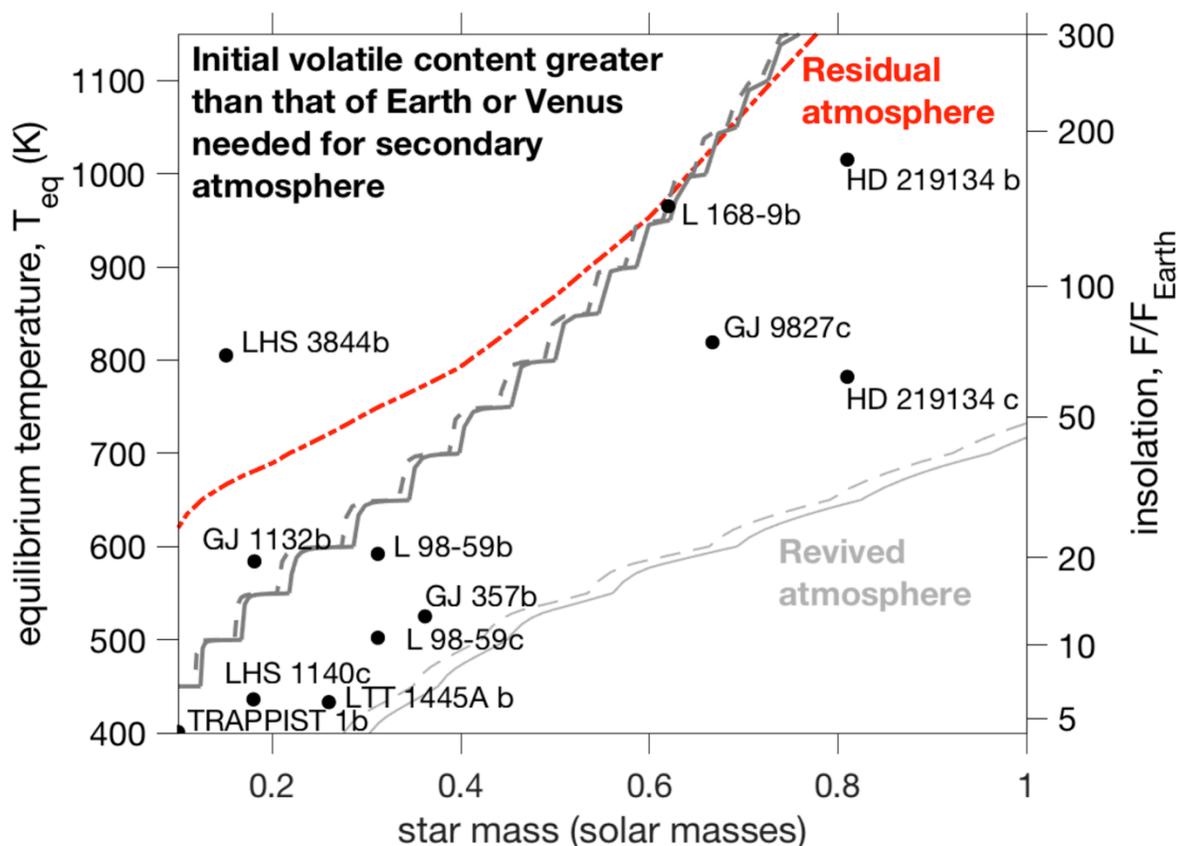

**Fig. S11.** Small planet atmosphere presence/absence diagram after 3.0 Gyr. As Fig. S10, but with a reduction in XUV flux by a factor of 10, or, equivalently, decreasing the efficiency of XUV-driven loss by a factor of 10, relative to the baseline based on the estimates of Selsis et al. 2007. This increases the range of exoplanets for which a volcanically supported atmosphere is possible. The red dash-dot line corresponds to the line of vanished atmospheres for planets that have all volatiles in the atmosphere initially. Worlds at hotter $T_{eq}$ than this line can retain an atmosphere if there is no $H_2$ initially and all high-molecular-weight volatiles are in the atmosphere initially. This line moves away from the star over time. The thick gray solid (plate tectonics) and dashed (stagnant lid) lines correspond to the line of atmospheric revival by volcanic outgassing for planets that lose all atmosphere during transition from a sub-Neptune to a super-Earth. The lines of atmospheric revival sweep towards the star over time because the rate of volcanic degassing falls off more slowly with time than does the star's XUV flux. This chart assumes initial volatile supply is independent of $F$ and similar to Earth and Venus. Increasing volatile supply will move thresholds to higher $F$. The thin gray lines show results for a reduction in the solubility of the high-molecular-weight volatile in the magma from $10^{-9}$ Pa$^{-1}$ to $10^{-11}$ Pa$^{-1}$. Note that because most of the planets that are shown are too small to have Earth-like composition and 6 $M_\oplus$, this plot is optimistic for atmospheric survival (higher planet mass favors atmosphere retention). For any individual planet, star-specific XUV-flux estimates, star age, and the planet's mass, should be combined to make a more accurate estimate than is possible using this overview diagram.

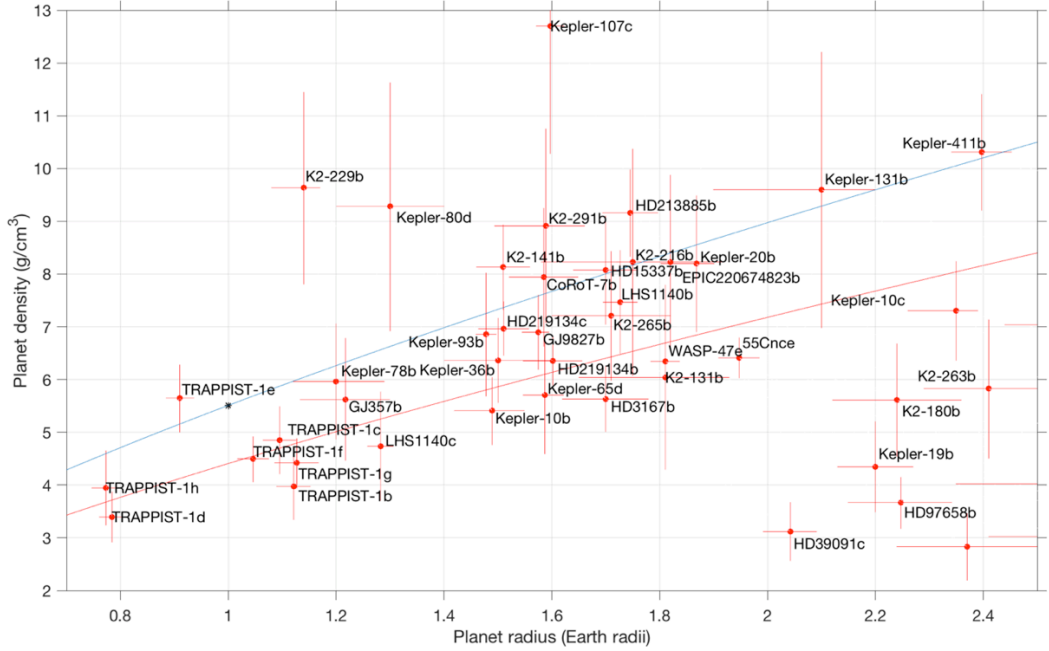

**Fig. S12.** Planet densities from Otegi et al. (2020). The blue line corresponds to density for Earth-like composition, assuming scaling of mass with planet radius as $(R/R_\oplus) = (M/M_\oplus)^{0.27}$. The red line corresponds to 80% of that density. Many super-Earth sized planets plot below the red line (for example HD 3167 b), indicating either a volatile envelope (atmosphere) or alternatively a very low (Mg+Fe)/Si ratio relative to that of the Earth. Super-Earths with the same or lower density as Earth (e.g. HD 3167 b) are excellent candidates for having elevated-molecular-weight secondary atmospheres. Planets plotting in the low-density, large radius clump in the bottom right are sub-Neptunes, retaining thick $H_2$-dominated atmospheres.

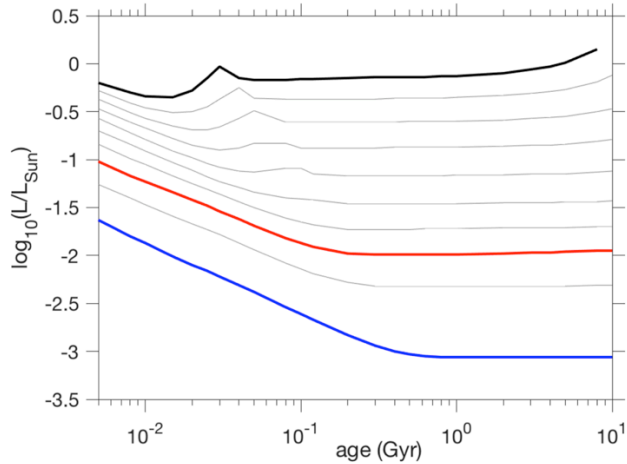

**Fig. S13.** Stellar luminosity, $L$, versus time from the model of Baraffe et al. (2015) for star masses ranging from 0.1 – 1.0 $M_\odot$, and normalized to the Sun's luminosity at the present-day ($L_\odot$). Contours drawn at intervals of 0.1 $M_\odot$. The blue line highlights 0.1 $M_\odot$. The red line highlights 0.3 $M_\odot$. The black line highlights 1.0 $M_\odot$.

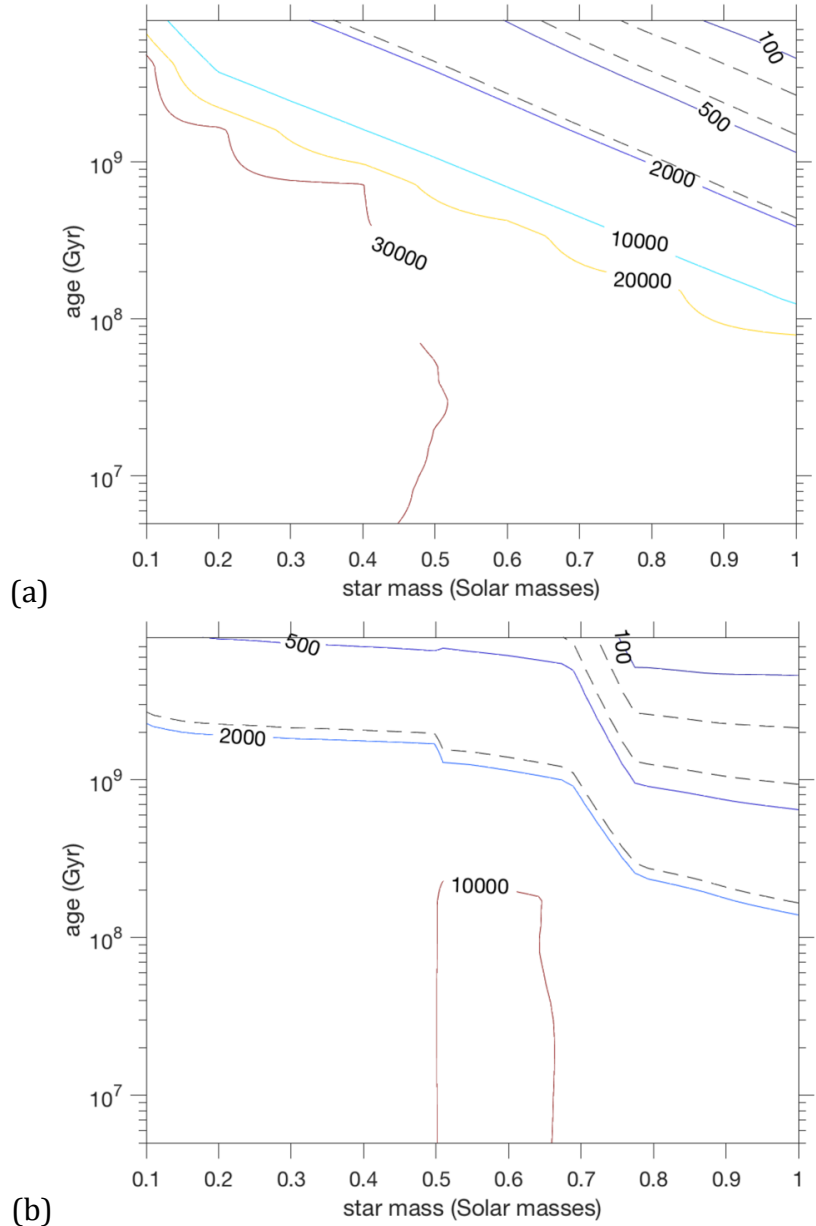

**Fig. S14.** $F_{XUV}$ at an insolation corresponding to a circular orbit 0.1 AU from today's Sun, in units normalized to the $F_{XUV}$ at Earth's orbit today. (a) for the model of Selsis et al. (2007), and (b) combining the fits of Jackson et al. (2012) and Guinan et al. (2016). The dashed lines highlight (from top to bottom) the threshold of $F_{XUV}$ below which all of the high-molecular-mass species is retained by the planet; the $F_{XUV}$ corresponding to a 50% reduction in the no-fractionation loss rate of the high-molecular-mass species; and the $F_{XUV}$ corresponding to escape of the high-molecular-mass species at 90% of the rate at which no loss would occur. Above the top dashed line, fractionation protects the constituents of the secondary atmosphere, and below the lowest dashed line, fractionation is much less important. Calculations are done assuming an atomic wind of H entraining a 1% mixing ratio of atoms of mass 15 Da, and planet mass 6 $M_\oplus$.

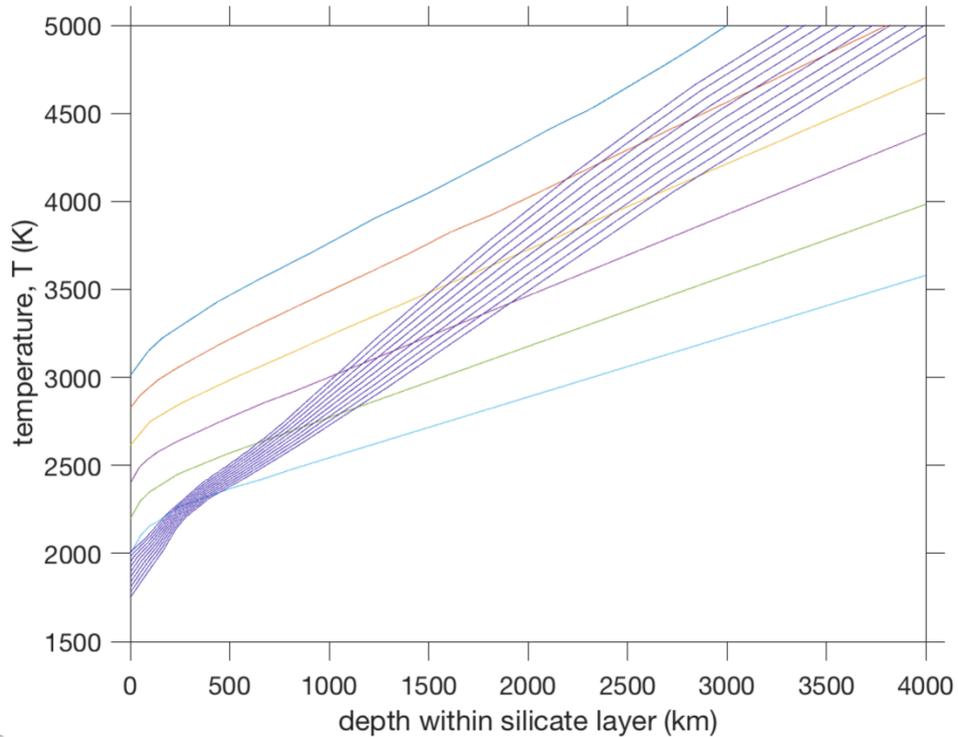

**Fig. S15.** Andrault et al. (2011) melt curves. Colored lines are magma adiabats (note that these are only physically meaningful for molten magma, i.e. for temperatures higher than the blue lines). Blue lines are melt fraction curves, where the lowest one is the solidus. We find that the effect of atmospheric overburden pressure on the solidus temperature is relatively small. X axis corresponds to depth within a hypothetical Earth-mass planet (the curves correspond to lines in P-T space, so they map to smaller depth on higher-gravity worlds).

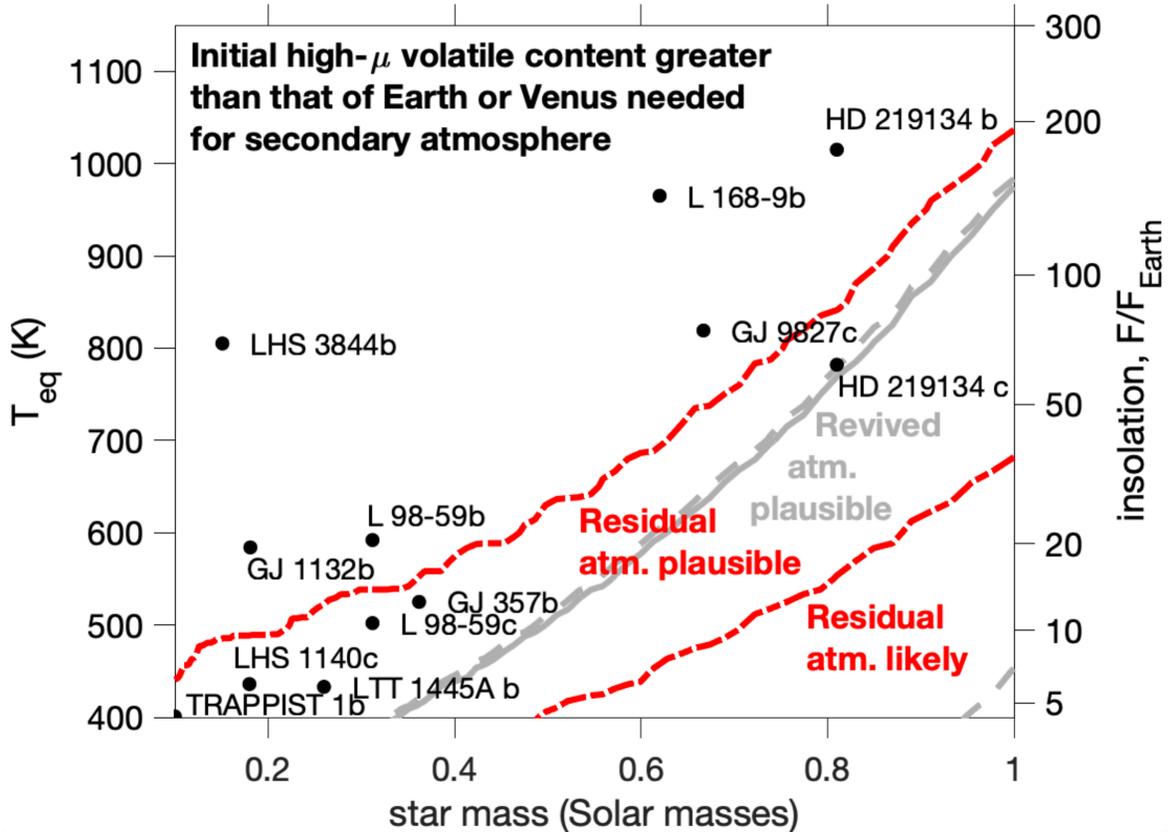

**Fig. S16.** As Fig. 6 (main text), substituting in an XUV-flux-vs.-time parameterization following Selsis et al. (2007). Secondary atmosphere presence/absence model output for 6 $M_\oplus$ (higher planet mass favors atmosphere retention). The dashed red lines show the lines of atmosphere retention after 3.0 Gyr for the case where all volatiles are in the atmosphere initially and there is no primary atmosphere; the 16th and 84th percentiles are shown, for varying XUV flux (by ±0.4 dex, 1σ; Loyd et al. 2020) relative to the baseline model following the results of Jackson et al. (2012) and Guinan et al. (2016) (SI Appendix, section 1a). These lines move away from the star over time. The gray lines show the 16th percentile for exhibiting an atmosphere after 3.0 Gyr for the case where volcanic outgassing rebuilds the atmosphere from a bare-rock state (the 84th percentile is at $T_{eq}$ < 400 K). The solid gray lines are for stagnant-lid tectonics and the dashed gray lines are for plate tectonics. The lines of atmospheric revival sweep towards the star over time because the rate of volcanic degassing falls off more slowly with time than does the star's XUV flux. In each case the atmosphere/no-atmosphere threshold is 1 bar. The black symbols show known planets that may be tested for atmospheres using JWST (Koll et al. 2019, Mansfield et al. 2019). For any individual planet, star-specific XUV-flux estimates, star age, and the planet's mass, should be combined to make a more accurate estimate than is possible using this overview diagram.

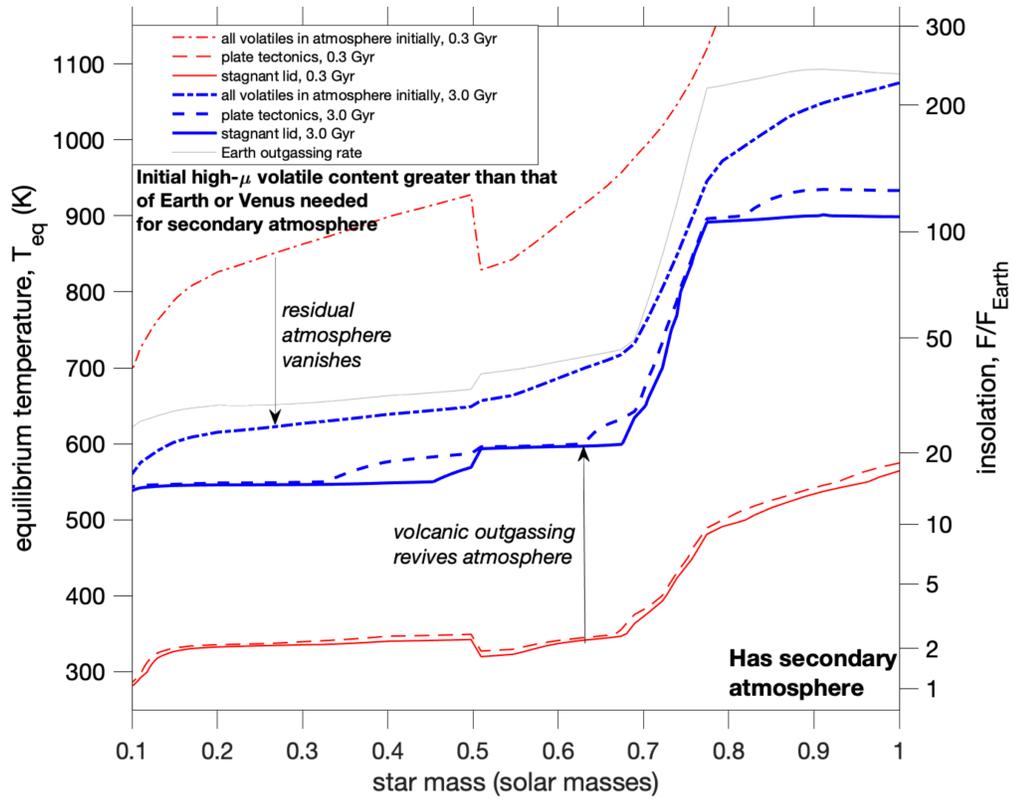

(a)

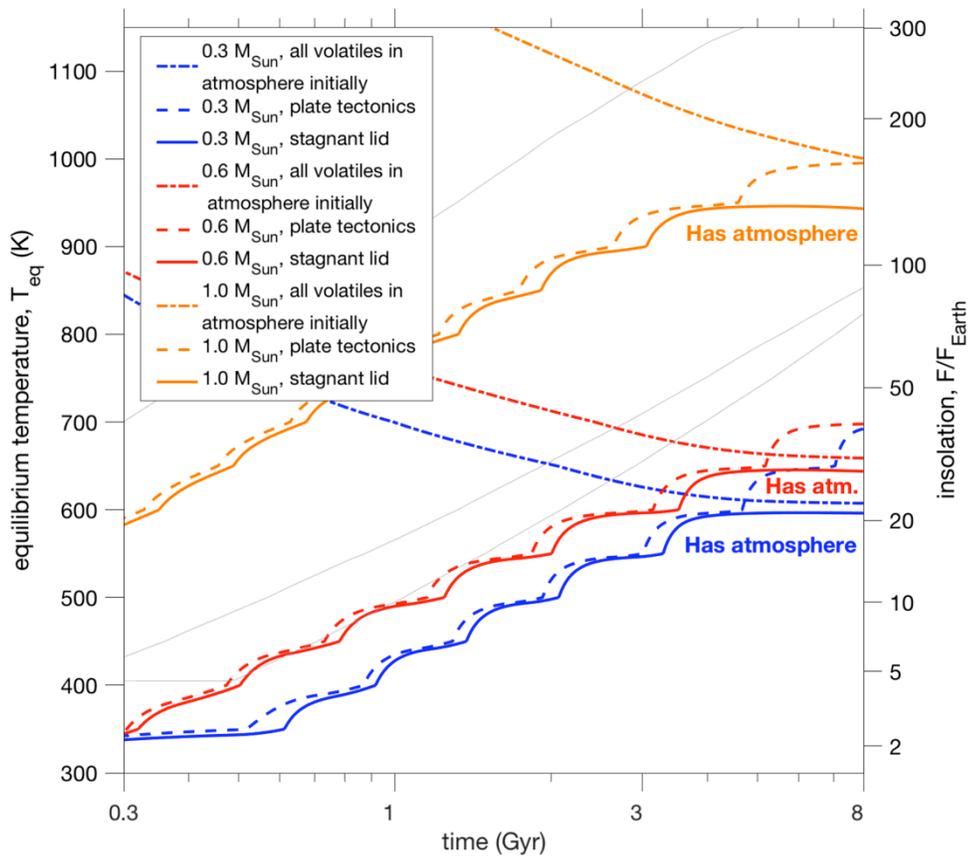

(b)

**Fig. S17. (a)** Secondary atmosphere presence/absence diagrams for 6 $M_\oplus$ (higher planet mass favors atmosphere retention). The dash-dot lines correspond to the line of atmosphere vanishing for planets that have all volatiles in the atmosphere initially. The solid and dashed lines correspond to the line of atmospheric revival by volcanic outgassing for planets that lose all atmosphere during transition from a sub-Neptune to a super-Earth. The line of revival sweeps toward the star over time because the rate of volcanic degassing falls off more slowly with time than does the star's XUV flux. The locations of these curves change depending on model assumptions. Increasing volatile supply will move thresholds to higher $L$. Allowing volatile supply to increase with decreasing $T_{eq}$, which is realistic, will steepen gradients with $L$. The gray line is for a constant Earth-scaled outgassing rate (72 bars/Gyr for 6 $M_\oplus$). **(b).** As (a), but showing changes over time. The gray lines are for 72 bars/Gyr outgassing at different star masses; from top to bottom, $\{0.3, 0.6, 1.0\}$ $M_\odot$.